\documentclass{aa} 
\usepackage{xcolor}

\usepackage[normalem]{ulem}

\usepackage{graphicx}
\usepackage{txfonts}
\begin{document}

   \title{Airy worlds or barren rocks? On the survivability of secondary atmospheres around the TRAPPIST-1 planets}
   \titlerunning{Airy worlds or barren rocks?}


   \author{Gwena\"el Van Looveren \inst{1}
          \and
          Manuel G\"udel \inst{1}
          \and
          Sudeshna Boro Saikia \inst{1}
          \and
          Kristina Kislyakova \inst{1}
          }

   \institute{Department of Astrophysics, University of Vienna,
              T\"urkenschanzstrasse 17, A-1180 Vienna\\
              \email{gwenael.van.looveren@univie.ac.at}
             }

   \date{Received September 27, 2023; accepted January 11, 2024}

 
  \abstract
   {The James Webb Space Telescope (JWST) is currently at the forefront of the search for atmospheres of exoplanets. However, the observation of atmospheres of Earth-like planets pushes the limits of the instruments, and often, multiple observations must be combined. As with most instruments, telescope time is unfortunately extremely limited. Over the course of cycle 1, approximately 100 hours have been dedicated to the TRAPPIST-1 planets. This system is therefore studied in unusually great detail. However, the first two sets of observations of the innermost two planets show that these planets most likely lack a thick atmosphere. The question therefore arises whether terrestrial planets around M stars have atmospheres or do not have atmospheres at all.}
   {We aim to determine the atmospheric survivability of the TRAPPIST-1 planets by modelling the response of the upper atmosphere to incoming stellar high-energy radiation. Through this case study, we also aim to learn more about rocky planet atmospheres in the habitable zone around low-mass M dwarfs.}
   {We simulated the upper atmospheres of the TRAPPIST-1 planets using the Kompot code, which is a self-consistent thermo-chemical code. Specifically, we studied the atmospheric mass loss due to Jeans escape induced by stellar high-energy radiation. This was achieved through a grid of models that account for the differences in planetary properties and irradiances of the TRAPPIST-1 planets, as well as different atmospheric properties. This grid allows for the explorations of the different factors influencing atmospheric loss.}
   {The present-day irradiance of the TRAPPIST-1 planets would lead to the loss of an Earth's atmosphere within just some hundreds of \textnormal{million years}. When we take into account the much more active early stages of a low-mass M dwarf, the planets undergo a period of even more extreme mass loss, regardless of planetary mass or atmospheric composition.}
   {The losses calculated in this work indicate that it is unlikely that any significant atmosphere could survive for any extended amount of time around any of the TRAPPIST-1 planets based on present-day irradiance levels. The assumptions used here allow us to generalise the results, and we conclude that the results tentatively indicate that this conclusion applies to all Earth-like planets in the habitable zones of low-mass M dwarfs.}


   \maketitle
%

\section{Introduction}

The search for habitable worlds has been a topic of growing interest in the past years. Especially with the launch of the James Webb Space Telescope (JWST), this search has gained renewed momentum, and multiple studies have investigated the suitability of JWST in the characterisation of terrestrial exoplanet atmospheres. The TRAPPIST-1 system is of particular interest due to its seven rocky planets \citep{Gillon2017}, four of which have masses comparable to the mass of Earth and lie within the habitable zone. Many studies (e.g. the review paper by \citealt{Turbet2020}, or more recent studies, e.g. \citealt{Turbet2022}, \citealt{Sergeev2022}, \citealt{Fauchez2022}) have proposed different types of atmospheric densities and compositions. These studies simulated how many observations would be needed for these different atmospheres to be observed and which signatures would be detected. In the light of the potential habitability of some of the TRAPPIST-1 planets, another key question emerges: whether these planets can even retain an atmosphere over a sufficiently long time to allow multi-cellular life to form and evolve. 

We investigate planets within the habitable zone of late-type stars. To do this, we take the TRAPPIST-1 system as an example. The circumstellar habitable zone is usually defined by the orbital distances at which a planet can lie for water to be potentially present on the surface in liquid form. The exact values for these distances are hard to define because many assumptions affect these calculations. The study by \citet{Kopparapu2013}, for example, discussed the habitable zones for various host stars. These calculations made use of the stellar flux incident on a planet in combination with climate models. The low mass of TRAPPIST-1 (0.0802 $\pm$ 0.0073~M$_{\odot}$, as reported by \citealt{Gillon2017}) would limit the habitable zone to between 0.03 to 0.06 au \citep{Kopparapu2013}. Several improvements have been made over the years to the models for the habitable zone (e.g. \citealt{Kopparapu2014}, \citealt{Kopparapu2016}). These improvements include the addition of factors such as the planetary mass and rotation rate. The stellar spectrum of these models takes into account the wavelength range, which spans from 230--4540~nm, comprising the brightness peak of the stars and the wavelengths reaching deepest into Earth's atmosphere. Photons at shorter wavelengths are absorbed in the upper layers of the atmosphere. However, and consequently, these uppermost layers are also the region of atmospheric escape. Hence, when the habitability of a planet is discussed, the atmosphere retainability of that planet should also be taken into account.

Using the Hubble Space Telescope, \citet{dewit2016, dewit2018} explored the possibility of hydrogen-dominated atmospheres for the TRAPPIST-1 planets. Their observations did not show any of the expected features, and cloud-free hydrogen-dominated atmospheres were ruled out for any of the planets. Recent JWST observations of TRAPPIST-1b and c (\citealt{Greene2023} and \citealt{Zieba2023}, respectively) seem to indicate that no thick atmospheres are present on either of these planets. \citet{Greene2023} presented photometric observations of the secondary eclipse of planet b. The day-side brightness temperature corresponds to a planet with little (less than a bar) to no atmosphere. Similar observations of TRAPPIST-1c by \citet{Zieba2023} ruled out CO$_2$ or O$_2$ dominated atmospheres down to 10 bar.

Several processes can cause the loss of an atmosphere. In the case of late-type main-sequence stars with convective envelopes, non-thermal losses such as charge-exchange escape or photochemical escape are highly important. These are the dominant loss mechanisms in the present-day Solar System (e.g. \citealt{Gronoff2020} and reference therein, \citealt{Lichtenegger2022}, \citealt{Scherf2021}). Since these mechanisms depend on the stellar wind parameters, they also become increasingly important for planets with small orbital distances. It is difficult to estimate non-thermal losses for exoplanets because they depend on the planetary magnetic field and the stellar equivalent of the solar wind, which are both extremely difficult to measure. Some examples of attempts to measure an exoplanet magnetic field and upper limits to stellar winds include \citet{Turner2021}, \citet{kislyakova2014}, and \citet{Fichtinger2017}, respectively, which further illustrate the complexity of these measurements. This means that estimates of non-thermal atmospheric losses are very uncertain and highly model dependent. Nevertheless, there have been attempts to model the stellar wind and associated losses of the TRAPPIST-1 system \citep{dong2017}.

Another category of atmospheric losses are thermal losses. This category includes hydrodynamic loss. This is an atmospheric loss that occurs when the atmosphere heats up to the point where the gas flows away hydro-dynamically. Several studies have parametrised the effect of hydrodynamic loss for Solar System planets under the early Sun (e.g. \citealt{tian2008}) and exoplanets in highly irradiated orbits (e.g. \citealt{owen2012}). Although this loss mechanism is not observed in the present-day Solar System, it is particularly important for hot Jupiters due to their proximity to their host star and the low molecular mass of their atmospheres. Observations of several planets (e.g. HD209458b \citealt{vidal-madjar2003}, or more recently, HAT-P-32b \citealt{zhang2023}) have shown that these planets can leave comet-like tails of hydrogen. A much more moderate form of thermal atmospheric loss is Jeans escape. This mechanism occurs when the kinetic energy of a particle leads to a velocity exceeding the escape velocity of the planet. The particles in the upper atmosphere gain much kinetic energy through the absorption of X-ray and ultraviolet (hereafter XUV) photons. Although this loss mechanism contributes only minimally to atmospheric escape in our present-day Solar System, it was much more prominent when the Sun was younger and more active \citep{johnstone2018, johnstone2021}. 
For many exoplanets, thermal losses can even become the dominant type of atmospheric escape. This is particularly true for exoplanet systems, in which planets  orbit their host star more closely and are thus more strongly irradiated, where host stars are much more active than our Sun. According to exoplanet.eu, about 46\% \footnote{2501/5420 as retrieved on 28/06/2023} of all detected exoplanets orbit their star more closely than Mercury orbits the Sun. The study of thermal escape is therefore of great interest. Additionally, Jeans escape is of particular interest because it requires fewer assumptions about the system and provides a lower limit to the mass-loss rate. 

The thermal mass-loss rate is often determined through the energy-limited escape rate \citep{Lammer2003}. This method determines the escape rate by estimating the temperature at the exobase based on the XUV flux received by the planet. This estimate takes several parameters of the planet into account, such as the gravity, the weight of the atmospheric components, and thermal conductivity, which are generalised for the whole atmosphere. Several other works have tweaked these parameters for specific cases (e.g. \citealt{Selsis2007} for H$_2$O) to better represent possible rocky worlds, but these parameters still represent generalised values. They can unfortunately not take into account effects such as (photo-)chemistry that changes the composition, or heating and cooling mechanisms. In order to self-consistently study the influence of thermal losses, atmosphere codes have been developed (e.g. \citealt{johnstone2018}, \citealt{Attia2021}, \citealt{nakayama2022}) that include several of these effects. In this work, we make us of the Kompot code \citep{johnstone2018}, which is discussed in section \ref{sec:code}, to study the atmospheric heating and Jeans escape of the Trappist-1 planets.

We study the thermal Jeans escape from the TRAPPIST-1 planets and its influence on their atmosphere and habitability. We opt to study these planets through a grid of simulated upper atmospheres, rather than simulations specifically tailored to a particular system, to account for a wider array of possible atmospheric compositions and to allow us to generalise our conclusions to other planets orbiting late-type M dwarfs.

In section \ref{sec:code} we explain the methods and models we used to simulate the atmospheric response to the incoming stellar XUV radiation for the TRAPPIST-1 planets. Next, we discuss the assumptions and implications of our models in section \ref{sec:ip}. In section \ref{sec:res} we present the results of the models, followed by the discussion and conclusions in sections \ref{sec:disc} and \ref{sec:conc}.

\section{Method}
\subsection{Model description}\label{sec:code}
Our work makes use of the upper atmospheric model known as the Kompot code, first described in \citet{johnstone2018}. Kompot calculates the 1D thermo-chemical structure of the upper atmosphere of a planet. 

The code was previously used to model various scenarios for Solar System planets. The work by \citet{johnstone2018} simulated the past Earth atmosphere going back to the late Archean. The loss rates for various atmosphere compositions under a younger Sun were used to constrain the minimum CO$_2$ required for a stable atmosphere. This study was expanded in \citet{johnstone2021} to further constrain CO$_2$ levels during the Archean and find that the Sun was likely born as a slow or moderately rotating star. The modelled atmospheres were also used to study non-thermal losses of the Archean Earth in \citet{Kislyakova2020}. Beyond Earth, the Kompot code was used to model losses of the early Martian atmosphere \citep{amerstorfer2017}. The code was expanded, as described in \citet{johnstone2020}, to include hydrodynamic losses. This is particularly important when we shift our attention towards exoplanets, in which irradiation is high enough for the atmosphere to become hydrodynamic \citep{owen2012,tu2015}.

To calculate the chemical structure, the code uses over 500 (photo)chemical reactions covering 63 species. These species include H, He, C, N, O, Ar, and Cl and combinations of these elements and their ions. For the full list of reactions, we refer to Appendix H of \citet{johnstone2018}.

To determine the thermal structure self-consistently, multiple different heating and cooling mechanisms are considered. The heating sources include X-ray, ultraviolet (UV), and infrared radiation; chemical reactions; and photo-electron heating. The radiation heating is handled by passing an input spectrum through each layer of the atmosphere following the laws of radiative transfer to take the changing chemical composition into account. The (photo)chemical reactions can also contribute to the heating according to their reaction rates and released energy. The photo-ionising reactions included in the network create electrons that further contribute to the heating of the atmosphere.

The cooling terms include, amongst others, CO$_2$ line cooling at 15 \textmu m, NO cooling at 5.3 \textmu m, and O-line cooling at 63 \textmu m and 147 \textmu m. These cooling mechanisms are calculated through analytical approximations. For a full list of equations and corresponding references, we refer to Section 2.5.2 of \citet{johnstone2018}.

Another important aspect of Kompot is that the code determines the exobase, the upper limit of the collisional atmosphere,
above which particles move ballistically (the start of "space", for our purposes). By allowing the code to dynamically change the upper limit of the simulated domain, we can model how an atmosphere becomes puffy when it heats up. Mathematically, we define this limit as the radius or altitude at which the Knudsen number equals unity. This means that the mean free path,
\begin{equation}
    \lambda_{\rm mfp} = \frac{1}{\sigma_{\rm Kn}n},
\end{equation}
equals the scale height,
\begin{equation}
    H = \frac{k_{\rm B} T}{m g}.
\end{equation}
In these equations, $\sigma_{\rm Kn}$ is a simplified collision cross section, $n$ is the number density, k$_{\rm B}$ is the Boltzmann constant, $m$ is the average molecular mass, and $g$ is the gravitational acceleration at the exobase. 

After the code setup using estimated initial conditions for the simulation parameters, we iterated the atmospheric model in time toward an equilibrium state. In each iteration, the code modelled a small time step in which the chemical and physical changes were applied sequentially. These time steps were then repeated until a stable solution was reached, that is, a self-consistent model of the upper atmosphere. This solution includes information on reaction rates, the thermodynamics of the system, the chemical composition, and the thermal structure. These last two points are most important in this work because they determine the Jeans escape. This was done following the method explained in section II.3 of \citet{bauer2004}. To determine the Jeans escape, we calculated the number of lost particles. This was done by first determining the thermal velocity,
\begin{equation}\label{eq:therm}
    v_{{\rm therm,}i} = \left(\frac{2k_{\rm B}T_{\rm exo}}{m_i}\right)^{1/2}.
\end{equation}
Here, k$_{\rm B}$ is the Boltzmann constant, $T_{\rm exo}$ is the temperature at the exobase, and $m_i$ is the mass of species $i$. Next, the Jeans escape parameter for species $i$ was defined as
\begin{equation}\label{eq:jeansparam}
    \lambda_{{\rm J,}i} = \frac{ G  M_{\rm p} m_i }{k_{\rm B} T_{\rm exo}  R_{\rm exo}}.
\end{equation}
Here, G is the gravitational constant, $M_{\rm p}$ is the planet mass, and $R_{\rm exo}$ is the exobase radius. Combining equations \ref{eq:therm} and \ref{eq:jeansparam}, we determined the effusion velocity of each species,
\begin{equation}
    F_{{\rm J,}i} = v_{{\rm therm,}i}  \frac{(1 + \lambda_{{\rm J,}i})e^{-\lambda_{{\rm J,}i}}}{2\sqrt{\pi}} .
\end{equation}
The effusion velocity integrated over the entire exobase gives us the mass-loss rate of the planet,
\begin{equation}\label{eq:massloss}
    \dot{M_i} = A  m_i  n_i  F_{{\rm J,}i}.
\end{equation}
Here, $n_i$ is the number density of species $i$ at the exobase, and $A$ is the surface area through which mass is lost. In this work, we use a 1D code, which implies certain limitations. Throughout this work, we model a column of atmosphere at an angle of 66$^{\circ}$  from the zenith, which on Earth is a decent representation of the average atmospheric structure \citep{johnstone2018}. In the case of slowly rotating or even tidally locked planets or atmospheres, there can be significant differences between the thermal structures on the day- and night-side. Measurements of the upper atmosphere of Venus, for example, show that the exobase temperature can vary over 100K \citep{limaye2017}. On the other hand, \citet{valeille2009} reported that 3D models showed that winds can redistribute energy in the upper atmosphere, causing the night-side to act as a heat sink. As the rotation periods of exoplanets or their atmospheres are usually not known, we assumed the extreme case in which particles are only lost on the day-side of the planet. In Eq. \ref{eq:massloss} this is represented by equating the loss surface $A$ to $2 \pi  R_{\rm exo}^2$. For planets with a fast-rotating atmosphere (e.g. Earth or Mars), this likely under-estimates the losses as the modelled column at a 66$^{\circ}$  angle from the zenith already represents a more moderate latitude.

Because of the dependence of the mass of the particle ($m_i$) and the number of particles present at the exobase ($n_i$), the mass-loss rate for an atmosphere needs to be averaged over all species. The species-specific mass-loss rates were weighted according to their mixing ratio at the exobase. We used a weighted average rather than a sum in order to prevent species that are not very abundant from skewing the total loss rate. For example, if an atmosphere composed of 99\% CO$_2$/1\%N$_2$ were to have a high N loss but low C loss, the nitrogen reservoir would deplete quickly, although the atmosphere as a whole would not be lost as fast. A consequence of using a weighted average is that we slightly underestimate the mass-loss rates, which we find preferable to overestimating the losses.

\subsection{Initial parameters}\label{sec:ip}
As shown in Sect. \ref{sec:code}, many factors influence the Jeans escape. Hence, we simulated a grid of different possible planets and atmospheres. This grid covers the planetary mass, the bulk composition of the atmosphere, and the received radiation. In this section, we discuss the parameter space covered by our simulations and the assumptions that these parameters imply. These parameters were chosen to cover most planets in the TRAPPIST-1 system, and they are summarised in Table \ref{tab:grid}.

The first boundary condition that needs to be set is the total number density at the lower boundary of the simulated domain. We set the lower boundary of the simulated domain to be around the pressure level of one millibar to avoid cloud-forming regions, which are not handled by our model. On Earth, this occurs at an average altitude of just below 50 km above sea-level. We assumed a number density of 1.73077$\times$ 10$^{16}$ cm$^{-3}$ and a temperature 258 K at this altitude, as calculated by the NRLMSIS-00 model for Earth in the Community Coordinated Model Centre \citep{picone2002}. As this is an empirical model, the values represent the parameters we could expect for an Earth-like planet with a surface pressure of 1 bar. 
However, this assumption is not a strict limitation to the models. Measurements of Venus, a planet with a surface pressure of 93 bar, show that the millibar level approximately lies an altitude of 85 km (\citealt{Limaye2018}), which is only 35 km higher than on Earth. When we calculated the gravitational acceleration at these different altitudes, the difference in escape velocity was smaller than 0.5\% for an Earth-sized planet. Therefore, we do not not expect the model results to vary greatly for planets with atmospheres more massive than Earth's for a given composition. The only difference would be that a more massive atmosphere would lead to a greater volatile reservoir, which would evidently require more time to be fully depleted.

The temperature at the base of the simulated domain was also taken from the NRLMSIS-00 model. Although a fixed temperature might result in some effects in the temperature profile that are not physically correct, a fixed temperature is needed for numerical stability at the lower boundary. However, the fixed-boundary condition only influences the cells closest to the boundary. At higher altitudes, and in particular, at the exobase, the temperature is dominated by other effects, such as heating from the star. We ran several tests models that showed that varying the base temperature in the range of 250 K to 550 K resulted in a spread of exobase temperatures smaller than 2\%. We here focus on planets with equilibrium temperatures below 400~K. It is therefore unlikely that the exact value of this base temperature significantly influences the loss rate.

We included C, N, O, and ions and molecules composed of the atoms. The mixing ratios of the chemical species were also fixed at the bottom boundary, which represents the mass reservoir of the lower atmosphere providing species that make up the bulk of the atmosphere. In this study, we varied the mixing ratios of this boundary layer from 1\%~N$_2$--99\%~CO$_2$ to 90\%~N$_2$--10\%~CO$_2$ to represent a set of likely (Venus- or Mars-like) atmospheric compositions. We focused on N$_2$ and CO$_2$ for several reasons. The first reason is that as summarised in \citet{Turbet2020}, observations with the Hubble Space Telescope exclude (cloud-free) hydrogen-dominated atmospheres. Further constraints from mass and radius measurements combined with data from the star favour atmospheres with a higher mean molecular weight. Another reason to focus on CO$_2$ is that as was shown in \citet{johnstone2021}, an increasing fraction of CO$_2$ can keep the exobase cool and to some degree prevents atmosphere loss under higher irradiation. Due to its higher molecular weight, CO$_2$ itself is also more resilient to atmospheric loss. Finally, we decided to discared H$_2$O dominated atmospheres as only the three innermost TRAPPIST-1 planets could have built up a steam atmosphere through the runaway greenhouse (see Section 2 of \citealt{Turbet2020} and references therein). However, \cite{johnstone2020} showed that atmospheres like this could not survive in the habitable zones of M dwarfs, let alone even closer to the host star.

The planetary mass and radius were also input parameters. According to \cite{Grimm2018}, the masses of the TRAPPIST-1 planets range from 0.3~M$_{\oplus}$ to 1.16~M$_{\oplus}$. Planets b and f have masses of 1.02 and 0.93~M$_{\oplus}$, respectively, making them similar to Earth, whereas planet e with its 0.77~M$_{\oplus}$ is closer in mass to Venus. Furthermore, planets c and g have masses of 1.16 and 1.15 M$_{\oplus}$, respectively. This apparently means that all of them are prime candidates for atmospheric observations. As the Jeans escape depends on the escape velocity, which in turn relies on the gravity of the planet, we modelled planets with 0.8, 1.0, and 1.2~M$_{\oplus}$ to cover the mass range of the larger TRAPPIST-1 planets. The radii of the low- and high-mass planets were scaled assuming the density of Earth (5.5~g~cm$^{-3}$), rather than other rocky bodies in our Solar System. This assumption led to higher escape velocities and thus to a more conservative estimate of the atmospheric loss.

\begin{table}
\caption{Overview of initial parameters of the simulated grid.}
\label{tab:grid}
\centering
\begin{tabular*}{\linewidth}{l p{0.4\linewidth} l}
\hline\hline
Parameter      &  Simulated values &  Unit\\
\hline
Irradiance  & 1, 2, 4, 6, 8, 10, 12, 14 & F$_{{\rm EUV,}\oplus}$ \\
Planet mass & 0.8, 1.0, 1.2             & M$_\oplus$       \\
Bulk composition & 10/90, 20/80, 40/60,  & \%CO$_2$/\%N$_2$     \\
& 60/40, 80/20, 90/10, &     \\
& 99/1 &     \\
\hline 
\end{tabular*}
\end{table}

\subsection{Stellar spectrum}
\label{sec:ip_spec}
We used the parameters of the TRAPPIST-1 system as a template for very low-mass M stars. The host star is of spectral type M8. These stars are fully convective and have a high X-ray to bolometric luminosity ratio, which they retain for longer periods of time than Sun-like stars (e.g. \citealt{Pizzolato2003, johnstone2021}). 

\begin{figure}
   \centering
   \includegraphics[width=\linewidth ]{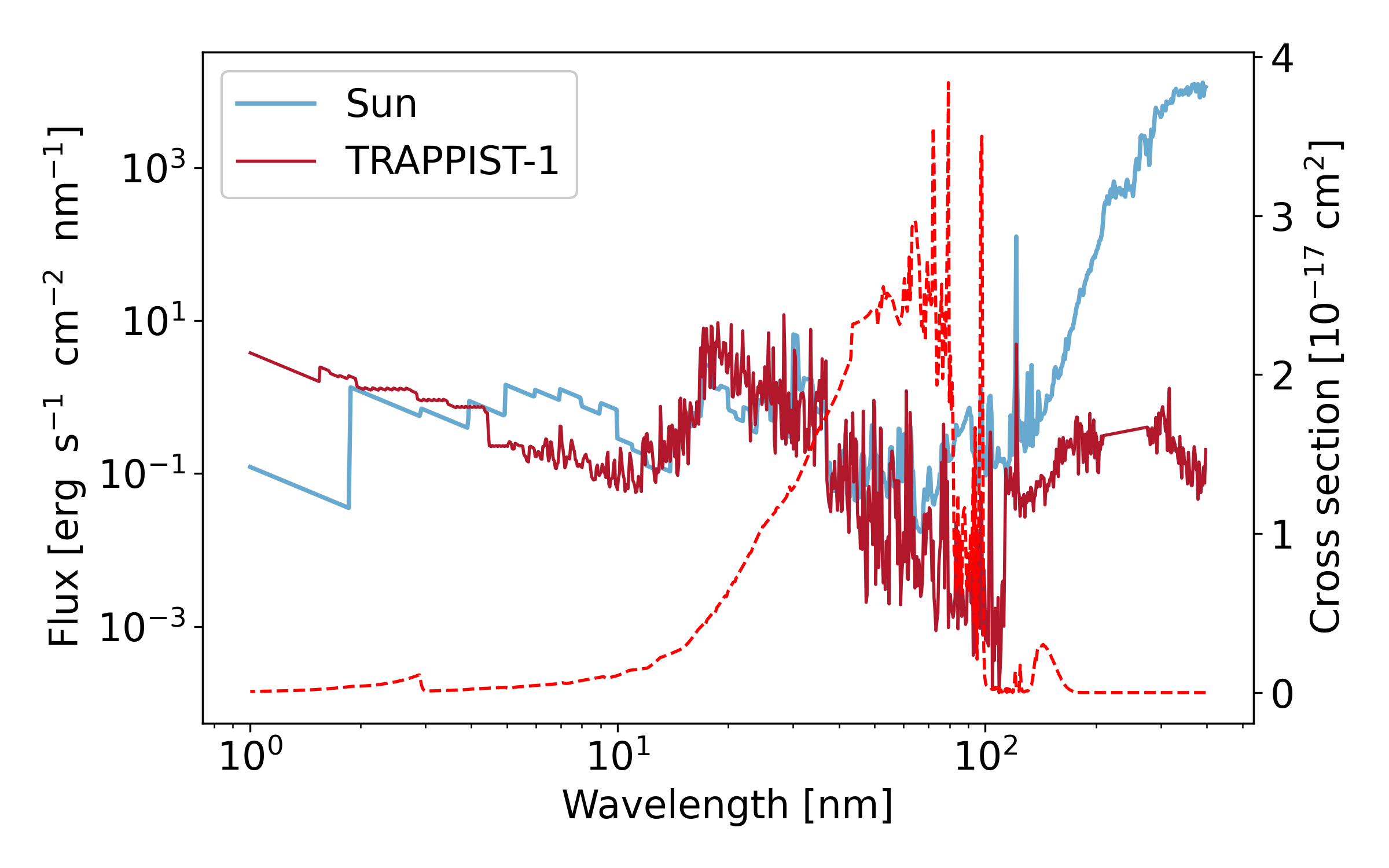}
   \caption{Comparison of the stellar spectra and absorption cross section. Left axis: Spectrum of the present-day Sun as seen at 1~au, plotted in blue  \citep{Claire2012}, and the spectrum of TRAPPIST-1 as received at 0.512~au, plotted in red \citep{Wilson2021}. Right axis: Abundance-weighted absorption cross section (dashed line) \citep{huebner2015} of an Earth-like atmosphere.}
    \label{fig:spectrum}%
    \end{figure}

Figure \ref{fig:spectrum} shows the present-day solar spectrum by \cite{Claire2012} in blue based on observations of the Sun and solar analogues. Throughout this work, we measure the irradiance in the extreme-UV (EUV) range, which we define as 10~nm (the boundary to the X-ray regime) to 121~nm (the Lyman-alpha line). We define one F$_{{\rm EUV,}\oplus}$ to be 4.77~erg s$^{-1}$~cm$^{-2}$, the integrated irradiance over this wavelength range of this spectrum. For each model, the entire XUV spectrum was used, but scaled according to this EUV flux.

The same plot also shows the modelled spectrum of TRAPPIST-1 from the Mega-MUSCLES survey \citep{Wilson2021} in red. In this figure, the spectrum has been scaled to 0.512~au, the distance at which a planet would receive one F$_{{\rm EUV,}\oplus}$. In order to scale this spectrum, we assumed a distance from Earth to TRAPPIST-1 of 12.43~pc, as was used in \citet{Wilson2021}. A comparison both of these spectra shows (as expected) that the irradiation in the near-ultraviolet is significantly lower for the later-type star. At short wavelengths, on the other hand, the TRAPPIST-1 spectrum is on par and in some spectral regions even slightly above the solar spectrum. This is in line with the observation of TRAPPIST-1 with XMM-Newton described in \citet{wheatley2017}. We calculated the X-ray luminosity (0.517 - 12.4~nm) from the Mega-MUSCLES spectrum to be about 3.4x10$^{26}$~erg s$^{-1}$, which corresponds to the low end of the luminosity range determined by \citet{wheatley2017}. 

The Mega-MUSCLES spectra were built using various models (XSPEC, optically
thin differential emission measure, semi-empirical models, and Phoenix; see \citealt{Wilson2021} for details). Whilst some parts of the spectra are based on observations, the EUV needed to be reconstructed due to interstellar absorption. The absorption cross sections of the dominant species in Earth's atmosphere, namely N$_2$, O$_2$, Ar, and CO$_2$, are large in the EUV. This is illustrated by the dashed red line in Figure \ref{fig:spectrum}, which was calculated based on the photo-ionisation and dissociation rates by \citet{huebner2015}. To limit the assumptions, we opted to take a conservative approach and use the solar spectrum by \cite{Claire2012}. We represent the irradiance received by the TRAPPIST-1 planets by scaling the solar spectrum. By minimising the number of assumptions, we aimed to keep our results more generally applicable to different stars. We discuss the implications of using different spectra in greater detail in Section \ref{sec:discstars}.

\section{Results}
\label{sec:res}
\subsection{Atmosphere structure} 

   \begin{figure}
   \centering
   \includegraphics[width=\linewidth ]{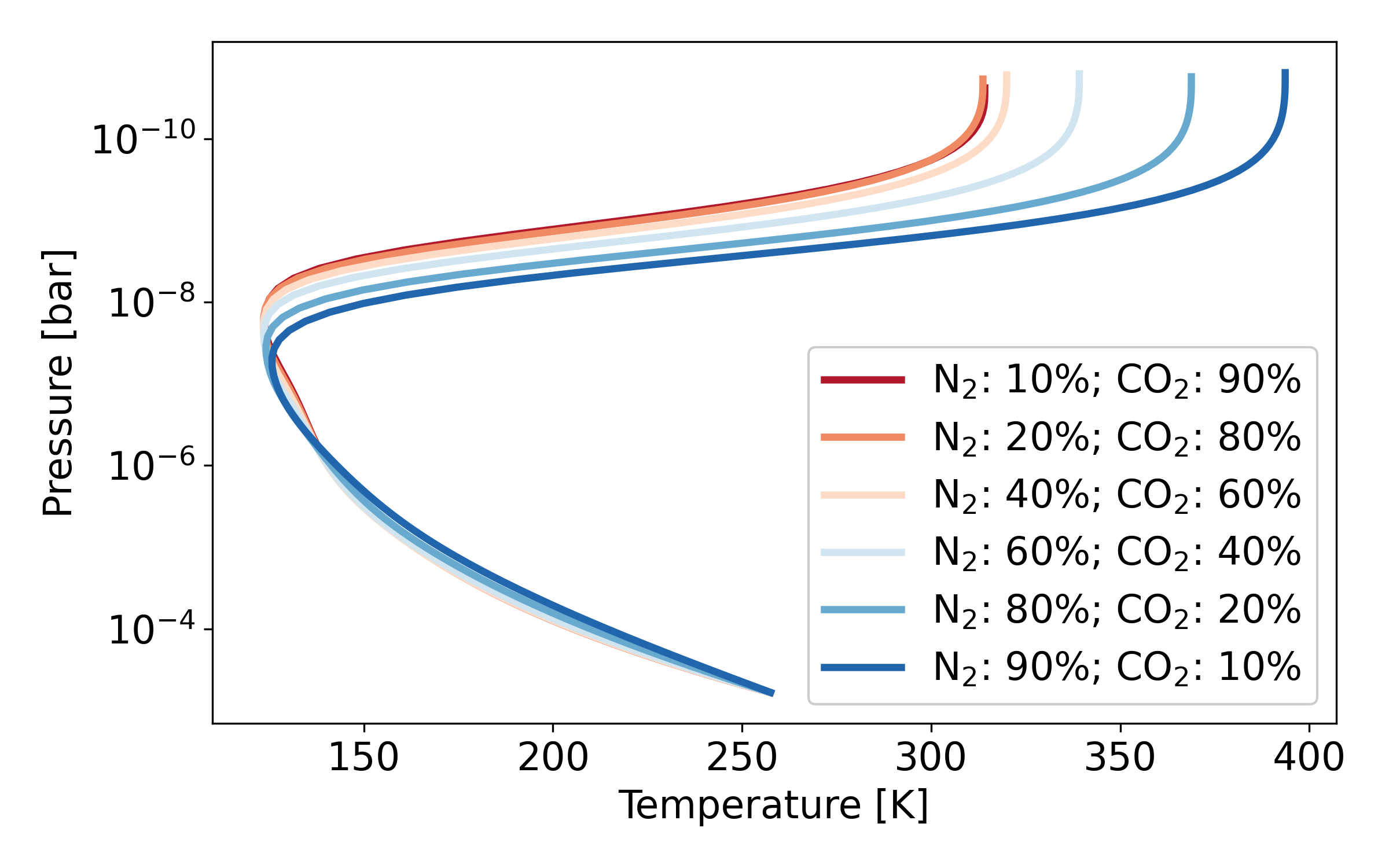}
   \caption{Pressure-temperature profiles for a 1 M$_{\oplus}$ planet receiving 1~F$_{{\rm EUV,}\oplus}$. The colours indicate the different bulk compositions.}
    \label{fig:pt}%
    \end{figure}

   \begin{figure*}
   \centering
   \includegraphics[width=0.9\linewidth ]{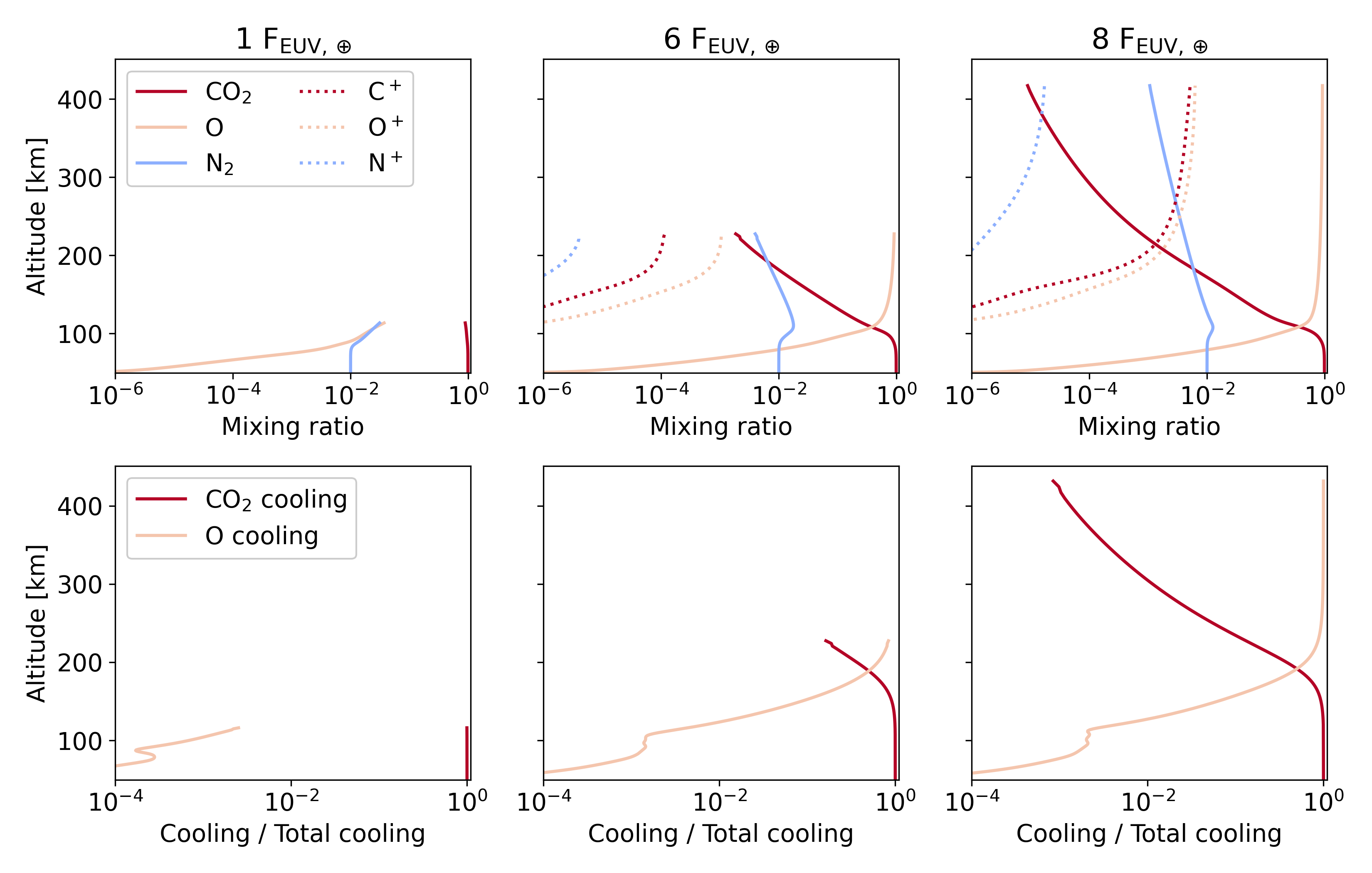}
   \caption{Mixng ratios and cooling contributions of Kompot models. Top row: Mixing ratios of several neutral species (indicated by solid lines) and ionised species (indicated by dotted lines) in the simulated domain. Bottom row: Contributions of different cooling mechanisms at different altitudes. Each column corresponds to a different levels of irradiation, namely 1, 6, and 8 F$_{\rm EUV,\oplus}$. All models presented in this figure are of a 1 M$_{\oplus}$ planet with an atmosphere composed of 1\%~N$_2$-99\%~CO$_2$.}
    \label{fig:mix}%
    \end{figure*}
    
The primary output of the Kompot code are pressure-temperature profiles and mixing-ratio profiles. Figure \ref{fig:pt} shows the pressure-temperature profiles for different atmospheric compositions for a planet of 1~M$_{\oplus}$ at 1~F$_{{\rm EUV,}\oplus}$. The figure shows that an increased CO$_2$ mixing ratio leads to a lower temperature near the top of the atmosphere, and in particular, at the exobase. The lower temperatures at similar pressures also indicate that the exobase altitudes are lower, which also contributes to a lower loss rate.

The top row of Figure \ref{fig:mix} shows examples of the vertical mixing ratio of a planet of 1~M$_{\oplus}$ with an atmospheric composition of 1\% N$_2$ and 99\% CO$_2$ at different irradiances. As each model contains dozens of species, we opted to only show some of the most abundant species and their associated ions. In the rest of this work, we use the abundance-weighted losses based on the uppermost layer of these profiles. These plots give insights into the processes taking place in our atmosphere. With an increasing amount of flux, the atmosphere expands to higher altitudes. In the low-irradiance scenario, the atmosphere remains closer to the surface and consists of mostly neutral species and molecules. In the high-irradiance scenarios, the uppermost layers of the atmosphere are significantly more ionised than the layers below, which is important for driving non-thermal losses, which are not the subject of our calculations, however. As discussed in section \ref{sec:code}, the total Jeans mass-loss rate depends on the masses and abundances of the escaping particles. In our examples, CO$_2$ photo-dissociates near the top of the atmosphere, creating atomic oxygen, amongst others. This is a much lighter particle that is lost more easily. Additionally, the creation of ions in this region provides an additional avenue of significant non-thermal escape driven by the stellar wind \citep{grasser2023}.

The second row of Figure \ref{fig:mix} shows the vertical distribution of CO$_2$ cooling and O cooling for the same simulations, taking the escape probability of the emitted photons into account. In these plots, the cooling mechanisms are normalised to the total cooling. There are two important points to note from this figure. Firstly, near the bottom of the simulated domain, where CO$_2$ is the dominant species, the cooling is almost entirely due to CO$_2$. In the 8 F$_{\rm EUV,\oplus}$ case, the top of the atmosphere is cooled primarily through O cooling as the upper most layers of the atmosphere become depleted of CO$_2$. However, in the lower layers, CO$_2$ cooling continues to contribute to the total cooling, even though the molecules become depleted at higher altitudes and irradiances. Secondly, the CO$_2$ cooling remains the dominant cooling mechanism up to an altitude of almost 200km, even though O becomes the dominant species at an altitude of 100km. This demonstrates the strength of CO$_2$ cooling.

\subsection{Composition variation} \label{sec:rescomp}
   \begin{figure*}
   \sidecaption
    \includegraphics[width=12cm]{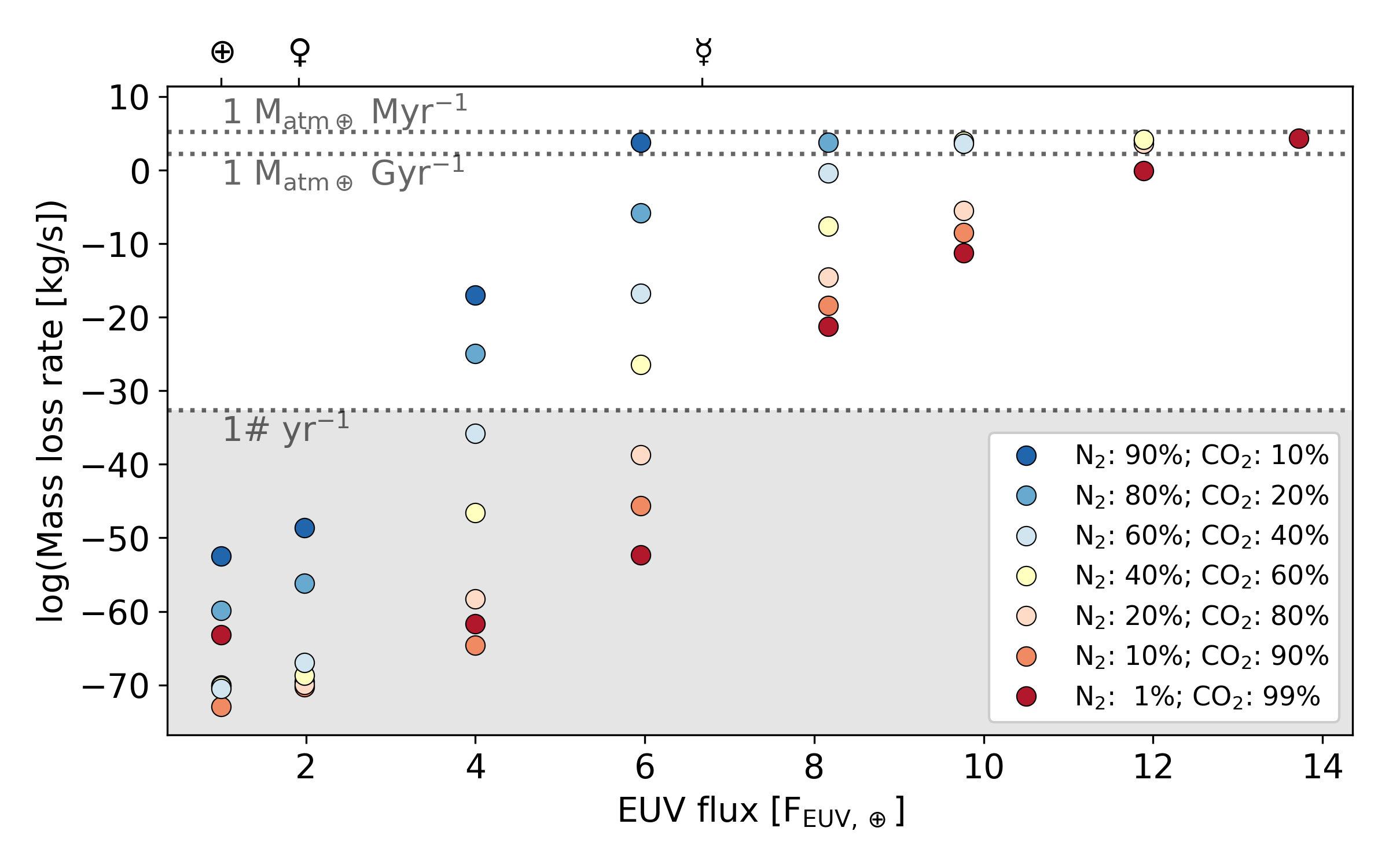}
        \caption{Abundance-weighted average Jeans escape for a 1~M$_{\oplus}$ planet, with bulk compositions coloured consistently with Fig. \ref{fig:pt}. The ticks on the top x-axis indicate the EUV flux received by the Solar System planets (given by their astronomical symbol) and the TRAPPIST-1 planets (indicated by letter). The horizontal lines indicate mass-loss rates in different units. The grey area indicates negligible loss rates.}
        \label{fig:m10_full}%
    \end{figure*}

In Figure \ref{fig:m10_full} we summarise the atmospheric mass-loss rates for a planet of one Earth mass. The y-axis shows the abundance-weighted mass-loss rate due to Jeans escape as a function of different irradiation levels and chemical compositions. The mass-loss rate in this figure covers over 70 orders of magnitude, but serves to illustrate that for most compositions and irradiation levels up to 6~F$_{{\rm XUV,}\oplus}$ Jeans escape is negligible. To differentiate realistic losses from purely numerical results, the lowest horizontal line and grey shaded area indicate a mass-loss rate lower than one single particle per year. The top two horizontal lines mark mass-loss rates that would result in the loss of the Earth's atmosphere (5.14x10$^{18}$~kg, \citealt{Trenberth2005}) in one billion and in one \textnormal{million years}, respectively, assuming a constant loss rate in time. For reference, the ticks on the top axis indicate the flux received by the inner Solar System planets (indicated by their astronomical symbols) up to an irradiance level of 14 F$_{{\rm EUV,}\oplus}$. In the TRAPPIST-1 system, this irradiance range corresponds to distances between 0.512 - 0.137~au. The TRAPPIST-1 planets are expected to receive approximately 68 (planet h), 120 (planet g), 177 (planet f), 306 (planet e), 529 (planet d), 1050 (planet c), and 1982 (planet b) F$_{{\rm EUV,}\oplus}$\footnote{These values are comparable to those estimated by \citet{wheatley2017}, although we caution against direct comparison as we used a different definition of the EUV range.}. These irradiances place all TRAPPIST-1 planets outside of the displayed range.

Figure \ref{fig:m10_full} shows that for all compositions, an increase in received flux leads to an increased loss rate. This is expected because as more energy is absorbed by the atmosphere, the temperature rises, and consequently, the loss rate increases. Atmospheres with a higher CO$_2$ content have a lower loss rate. This can be attributed to the cooling through the 15\textmu~ line, which is also reflected in Figure \ref{fig:pt}. This trend agrees with previous work by \citet{johnstone2018} and \citet{johnstone2021}, where various atmospheres were simulated for planets around a young Sun. 

At 1 F$_{{\rm EUV,}\oplus}$ the 99\% CO$_2$ model seemingly does not follow the trend of increasing loss with increasing irradiation. We emphasise again that these mass-loss rates correspond to orders of magnitude less than a single electron over the age of the Universe. Due to the discretisation of the atmosphere, the simulations can fluctuate around a value between two cells, which can result in seemingly large differences in the mass-loss rates, depending on the final state in which the model settles. All mass-loss rates in the grey shaded area of all figures are therefore purely numerical output and should be considered zero.

At six times F$_{{\rm EUV,}\oplus}$ the N$_2$ dominated atmosphere (90\% N$_2$) reaches a mass-loss rate resulting in the loss of the entire Earth's atmosphere on a timescale of 10 Myr. At higher fluxes, this model reaches temperatures that fall outside of the domain in which some of the analytic expressions for heat exchange used in the code are valid. At 8 F$_{{\rm EUV,}\oplus}$ the atmosphere with 80\% N$_2$ reaches the same mass-loss rate, and at 12 F$_{{\rm EUV,}\oplus}$, even the CO$_2$ dominated atmosphere (90\% CO$_2$) is lost within a few \textnormal{million years}.

   \begin{figure}
   \centering
   \includegraphics[width=\linewidth ]{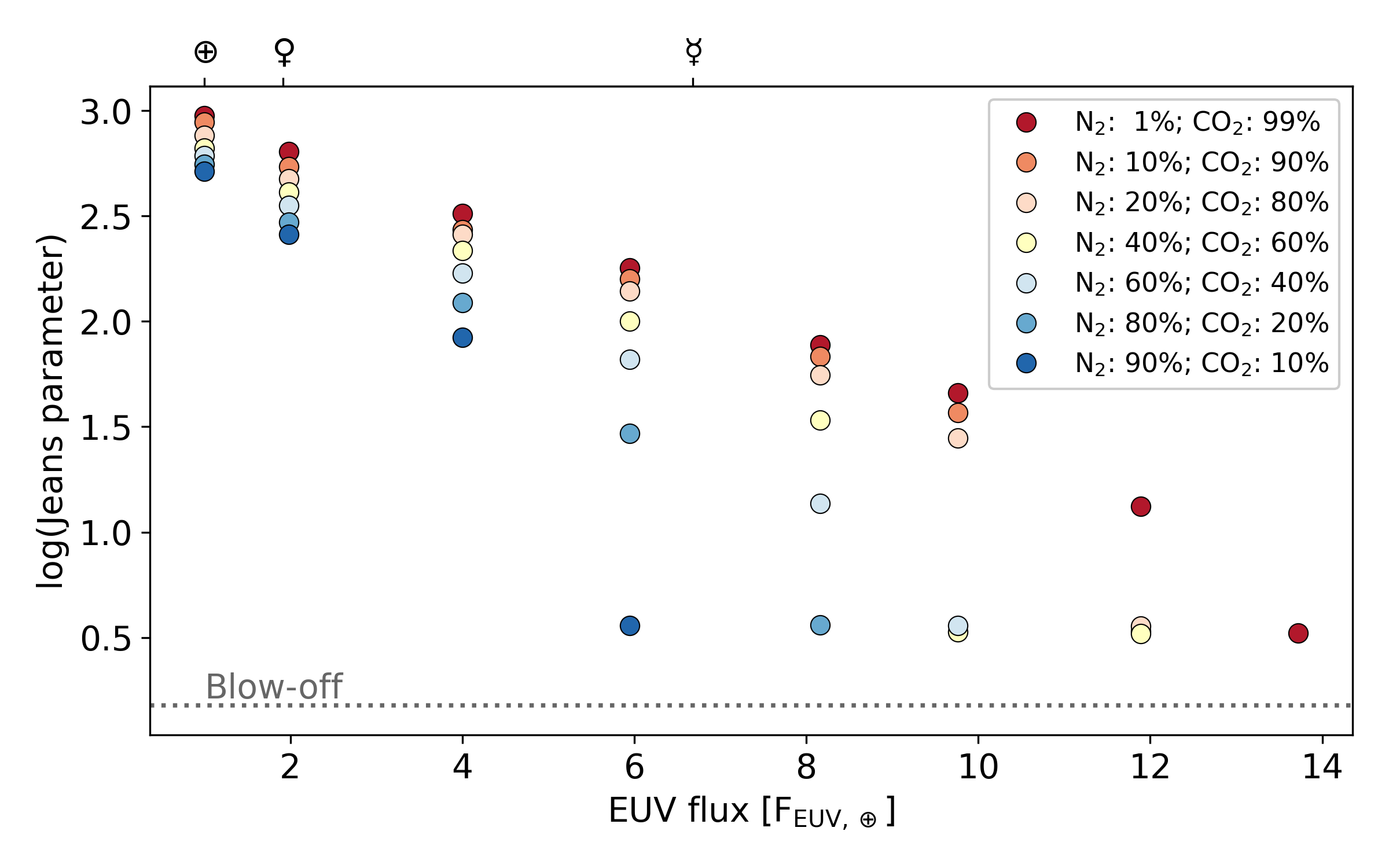}
        \caption{Jeans parameter for a 1~M$_{\oplus}$ planet. The bulk compositions are coloured consistently with Fig. \ref{fig:pt}. The ticks on the top x-axis indicate the EUV flux received by the Solar System planets (given by their astronomical symbols). The horizontal line indicates the blow-off limit.}
        \label{fig:m10_jeans}%
    \end{figure}

In Figure \ref{fig:m10_jeans} we show the same models as in Figure \ref{fig:m10_full}, but we use the Jeans parameter as defined in Eq. \ref{eq:jeansparam}. This figure serves as an easy comparison to previous research as many studies on Jeans escape used this parameter rather than the mass-loss rate. The horizontal line in this figure indicates a Jeans parameter of 1.5 (or 0.4 on the logarithmic axis of Fig. \ref{fig:m10_jeans}), which corresponds to the blow-off condition identified by \citet{opik1963}. We note that none of the models reaches this limit, even though the corresponding mass-loss rates are very high. This condition was traditionally considered the transition point between Jeans escape and hydrodynamic escape, although it was discussed by \citet{tian2008} that an atmosphere reaches the hydrodynamic flow regime before it reaches this blow-off condition, which means that other effects become important. For the remaining figures throughout this study, we therefore use the mass-loss rate rather than the Jeans parameter.

\subsection{Mass variation}
\label{sec:resmass}

   \begin{figure}
   \centering
   \includegraphics[width=\linewidth ]{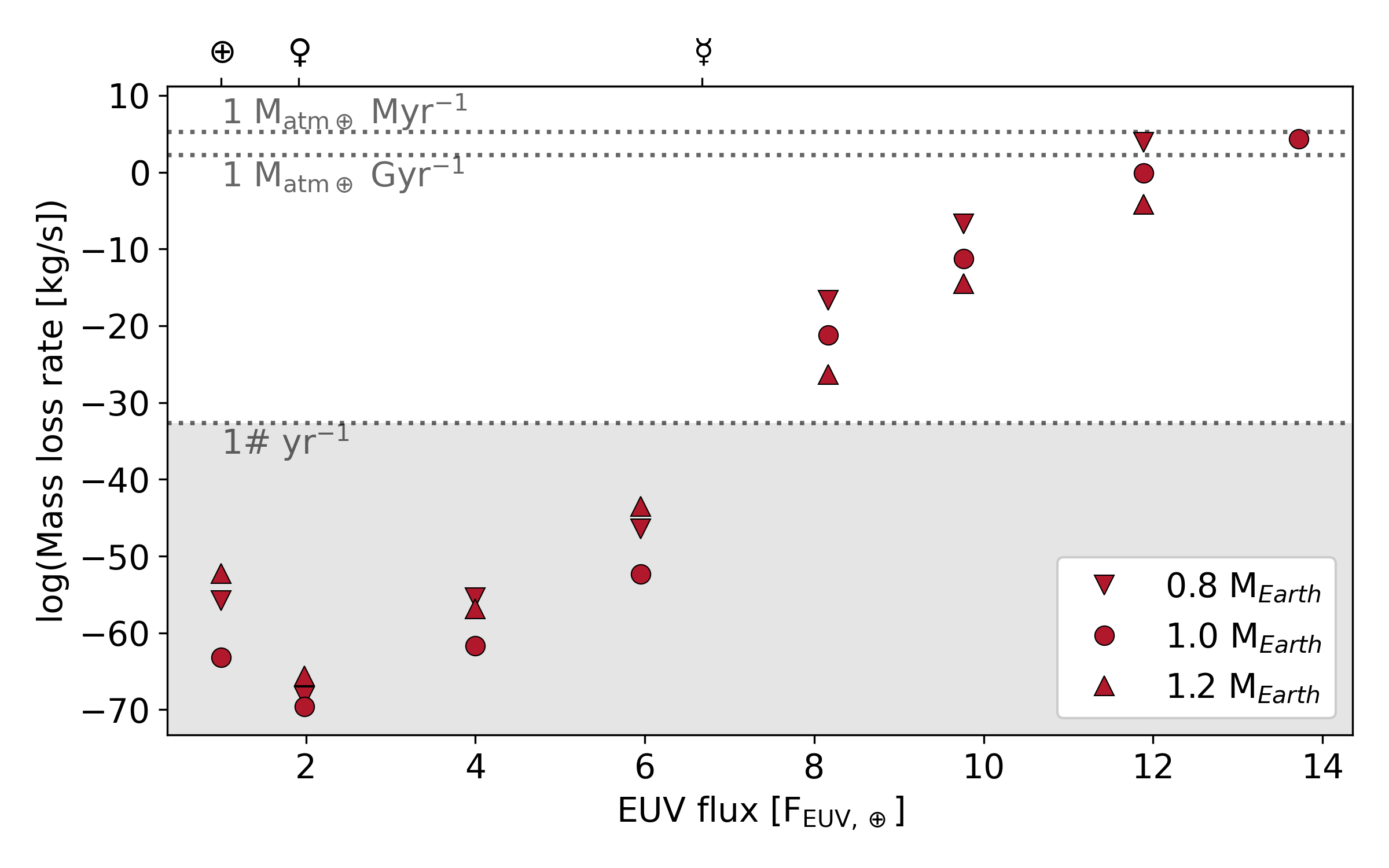}
   \caption{Abundance-weighted average Jeans escape for a planet with a 1\%~N$_2$-99\%~CO$_2$ (indicated by the same colour as in Fig. \ref{fig:m10_full}). The markers indicate different planetary masses, which are consistent with Fig. \ref{fig:m10_full}. The ticks on the top x-axis indicate the EUV flux received by the Solar System planets (given by their astronomical symbols). The horizontal lines indicate mass-loss rates in different units. The grey area indicates negligible loss rates.}
    \label{fig:massvar}%
    \end{figure}

As discussed in Section~\ref{sec:code}, the mass of the planet has an impact on the Jeans escape. Figure \ref{fig:massvar} shows the mass-loss rates for an atmosphere composed of 1\% N$_2$ and 99\% CO$_2$. As seen in section \ref{sec:rescomp}, this is the composition with the lowest mass-loss rates. The figure shows that the lowest-mass planet has the highest escape rate and the highest-mass planet has the lowest escape rate. The escape rate is inversely proportional to the escape velocity. Because we scaled the different planetary radii using the Earth's density, the escape velocity at the base can be rewritten as
\begin{equation}
    v_e = M_{\rm p}^{1/3} \left(\frac{32 \pi G^3 \rho}{3}\right)^{1/6}.
\end{equation}
This equation shows that we expect an escape velocity at 93\% and 106\% of the Earth-mass planet value for the low- (0.8~M$_{\oplus}$) and high-mass (1.2~M$_{\oplus}$) planet, respectively. However, Figure \ref{fig:massvar} shows mass-loss rates that are considerably higher than what was estimated using the escape velocity. This can be attributed to the fact that the exobase of the atmosphere is determined by the scale height, which also depends on the planetary gravity. The atmosphere of a lower-mass planet will expand more easily, causing the complete loss of the atmosphere at lower fluxes. On the other hand, the higher mass reduces the mass-loss rate to some degree, but not enough to prevent the loss of the atmosphere at higher irradiation levels.

For the sake of completeness, we add figures showing the averaged mass-loss rate for all modelled compositions for the lower- and higher-mass planet in App. \ref{app1}.


\section{Discussion}
\label{sec:disc}
\subsection{Current TRAPPIST-1 system}
The results shown in Section \ref{sec:rescomp} demonstrate that a planet receiving more than 6 F$_{{\rm EUV,}\oplus}$ is likely to experience significant thermal losses. This assumes that no major unknown sources replenish the atmosphere, which we discuss in section Sec. \ref{sec:evol}. In section \ref{sec:resmass} we showed that even the largest modelled planet with a CO$_2$ dominated atmosphere would experience extreme mass-loss rates if it received more than 14~F$_{{\rm EUV,}\oplus}$; this includes all of the TRAPPIST-1 planets. Although only planets d, c, and b are too close to their host star to harbour liquid surface water, our results indicate that none of the planets should be able to hold on to a N$_2$-CO$_2$ atmosphere. This is compatible with the initial results by \citet{Greene2023} and \citet{Zieba2023}, who excluded dense atmospheres around TRAPPIST-1 b and c, although more observations are required to assess the potential presence of thin atmospheres. \citet{wheatley2017} estimated the mass-loss rates of these inner planets using the energy-limited approximation, and the authors concluded that these planets are probably stripped of any atmosphere on a timescale of some \textnormal{billion years}. The authors assumed a ratio of F$_{\rm EUV}$/F$_{\rm X}$ equal to 1.31 or 1.78. When we calculate the same ratio based on the MUSCLES spectrum of TRAPPIST-1, we find F$_{\rm EUV}$/F$_{\rm X}$ equal to 9.9. For the solar spectrum that we used in our models, this ratio is equal to 5.1. This difference explains why our mass-loss rates are seemingly higher than the energy-limited mass-loss rates estimated by \citet{wheatley2017}.

The planets in the habitable zone, TRAPPIST-1 e, f, and g, are of central interest. Due to their close-in location, these planets are also unlikely to have retained an atmosphere (see Figure \ref{fig:massvar}) given the estimated age of the system of 7.46$^{+2.01}_{-2.10}$ Gyr \citep{fleming2020} or 7.6$\pm$2.2 Gyr \citep{Burgasser2017}. Even when we assume the lowest age estimate and an atmosphere 100 times the mass of Earth's atmosphere, the losses of the order of 1~bar per 5 Myr that we find at 14 F$_{\rm EUV,\oplus}$ suggest that an observable atmosphere is unlikely for any of these planets.

\subsection{Atmospheric loss on evolutionary timescales}\label{sec:evol}
Figure \ref{fig:m10_full} shows that an Earth-mass planet would only experience relatively low mass-loss rates or even insignificant losses for certain compositions for irradiances lower than 10 F$_{\rm EUV,\oplus}$. For the TRAPPIST-1 system, this is beyond 0.16~au. Even TRAPPIST-1 g, with its estimated mass of 1.15 M$_{\oplus}$ \citep{Grimm2018}, which appears to be the most interesting candidate for studies of a habitable atmosphere, falls well inside this limiting radius.

Furthermore, it is likely that the TRAPPIST-1 planets experienced stronger atmospheric loss in their early history because the stellar activity level also evolved. It most likely declined over time due to stellar spin-down. The importance of studying atmospheres at different stages of the host star evolution is further shown by the case of the early Earth, whose atmosphere was subject to very different and changing conditions because of the considerably evolving high-energy radiation output of the Sun as it has been spinning down (e.g. \citealt{johnstone2021}).

After their formation, mid- to late-M dwarf stars evolve very slowly to the phase of stable hydrogen burning. A star like TRAPPIST-1 takes about 1--2~Gyr to settle on the main sequence \citep{ramirez2014}. During this period, the X-ray activity usually stays at a saturated level of $\sim$10$^{-3}$ times the bolometric luminosity. Because M dwarfs move vertically down the Hayashi track in the Hertzsprung-Russell diagram, the initial bolometric and X-ray luminosities were therefore higher by between one and two orders of magnitude than after reaching the main sequence. Planets in stable orbits will therefore have been subject to much higher irradiation levels than after the host star reached the main sequence even when they are still at saturation.

Consequently, the habitable zone also shrank substantially during the first \textnormal{billion years} in the life of such a planetary system; the bolometric luminosity of TRAPPIST-1 decreased by a factor of 40 from an age of 10~Myr, when terrestrial planets have supposedly reached their near-final mass (e.g. \citealt{lammer2020} for the Solar System) to $\sim$1--2~Gyr when the star arrived on the main sequence \citep{fleming2020}. This means that the orbital distance of the habitable zone shrank by a factor of $\sqrt{40} = 6.3$. \citet{bolmont2017} estimated the inner edge of the habitable zone of a 0.08~M$_{\odot}$ star to move from $\sim$0.08~au at an age of 10~Myr to $\sim$0.012~au at an age of 2~Gyr on the main sequence (their Fig.~1b).
This implies that all known TRAPPIST-1 planets (with semi-major axes of $\sim$0.01--0.06~au) were far inside the habitable zone for the first tens to hundreds of \textnormal{million years} when secondary atmospheres are typically assumed to build up. It is likely that potential water oceans would have evaporated, and the TRAPPIST-1 planets would have followed the fate of Venus \citep{ramirez2014}. We note that the loss rates of hydrogen produced by photo-dissociation of evaporated H$_2$O are much higher than the loss rates of the heavy atoms reported in our work.

Once on the main sequence, mid- to late-M dwarfs continue to be appreciably active. This is a consequence of their deep convection zones, combined with sufficiently fast rotation rates as ingredients for efficient magnetic dynamo action \citep{johnstone2021a}. The rotation rates of stars with masses $<0.2~M_{\odot}$ are typically much higher than the solar rotation rate for ages of $\sim1$~Gyr.
More importantly, the convective turnover time is much longer for a 0.1M$_{\odot}$ star than for a 1M$_{\odot}$ star even at 2~Gyr, implying that the former is still at the X-ray saturation level, as mentioned above. The subsequent decline of its X-ray activity level due to spin-down is very slow and is only about one order of magnitude in several \textnormal{billion years}, in contrast to a Sun-like main-sequence star that decays in X-rays by 2--3 orders of magnitude on the same timescale while on the main sequence \citep{johnstone2021a}. The XUV flux in the habitable zone remains correspondingly high for several \textnormal{billion years} on the main sequence.

As mentioned before, this argument assumes that the secondary atmosphere was present soon after the planets had formed. The majority of a secondary atmosphere would form during the primary outgassing, when the magma ocean solidifies, which occurs soon after the planet formation \citep{vanhoolst2019}. The timescales on which this occurs depend on many factors. To name just two factors that can greatly influence the atmospheric formation timescale, there is a dependence on how the magma ocean solidifies, and how a primary atmosphere transitions into a secondary atmosphere \citep{lichtenberg2023}. The primary outgassing is likely limited to the first \textnormal{billion years} after formation as a result of the thermal budget of the planet. Secondary outgassing might still take place well beyond this point in time, although estimating the amount of released volatiles or the outgassing rate is exceedingly complicated because it relies on factors such as the presence of plate tectonics, the planetary mass, and the composition of the planetary interior \citep{vanhoolst2019}. Volcanic outgassing could contribute several to tens of bars of CO$_2$ to an atmosphere, but for planets of approximately one Earth mass, the majority of the stored gasses are depleted within 1 Gyr \citep{dorn2018}. After this amount of time, the X-ray activity of late-type M dwarfs usually remains saturated, meaning that the atmospheric loss rates will remain high for the considered planets. Other methods of volatile delivery, such as comet impacts, could also replenish the atmosphere to some degree, but this would be equally difficult to quantify for exoplanets and most likely only contribute small, undetectable amounts. 

\subsection{Other loss mechanisms}
We also point out that we only considered atmospheric losses through Jeans escape so far. According to \citet{tian2008}, an Earth-like atmosphere would become hydrodynamic if it were exposed to an irradiation above five times the Earth's current irradiation. This would mean that the comparatively low Jeans escape turns into a hydrodynamic flow, resulting in even greater losses than those calculated here. Furthermore, non-thermal effects, such as charge exchange, will continue to increase the loss rate. Due to the proximity of the planets to their host star, these losses are likely substantial as well. As demonstrated in \citet{dong2017}, the ion escape can lead to the loss of an Earth-like atmosphere in the order of 0.1 - 10 Gyr (depending on the planet), which is comparable to the thermal losses calculated in this work.

The mass-loss rates presented in this work are therefore lower limits. Although there are some further mechanisms that could lower the mass-loss rate, these are unlikely to make a significant difference given the magnitude of the losses. For example, the cooling mechanisms included in the Kompot code rely on analytical approximations and do not include an exhaustive list of significantly cooling lines. In a similar study, \citet{nakayama2022} used a more extended treatment of cooling through infrared emission lines of C and O. They found stronger cooling and consequently smaller thermal losses based on this, but they also identified stronger ionisation, and consequently, stronger recombination in the upper atmospheric layers than we do. Future studies need a careful reconsideration of the balance between ionisation, recombination, heating, and cooling.

Another factor to consider would be the presence of a magnetic field. Planetary magnetic fields have long been considered to protect atmospheres against interaction with the stellar wind (e.g. \citealt{khodachenko2012}). However, a magnetic field adds escape through polar outflow and cusp escape. \cite{Gunell2018} showed that due to these additional loss mechanisms, planets with magnetic fields can have similar or even larger losses than planets without magnetic fields. These losses are primarily driven by ions in the upper atmosphere, which are much more present in the more irradiated models presented here due to photoionisation. This was demonstrated in \citet{grasser2023} for the Archean Earth, which is also a highly irradiated planet. This once again supports our view that the losses determined here should be seen as lower limits.

\subsection{Role of the stellar spectrum}
\label{sec:discstars}
Stellar XUV radiaton is the key driver of planetary atmospheric loss. Hence, it is important to consider these shorter wavelengths in the search for possibly habitable atmospheres around exoplanets. As a result of the interstellar absorption in the EUV wavelength range, the full XUV spectrum has to be reconstructed, as mentioned in Section \ref{sec:ip_spec}. When we compare the reconstructed spectrum of TRAPPIST-1 against the observed solar spectrum, we note a few key differences, some of which are due to differences in stellar properties, and some are due to assumptions in the applied reconstruction technique. In the X-ray section of the spectra, the flux of the TRAPPIST-1 spectrum around 1~nm is higher by almost two orders of magnitude than the solar spectrum. However, around 10~nm, the solar spectrum becomes higher by about one order of magnitude. Another difference is visible around 100~nm. This wavelength range is not dominated by stellar photospheric contributions (which are very different between M- and G-type stars), but by chromospheric, transition region, and coronal contributions, that is, by stellar activity.

\begin{figure*}
   \centering
   \includegraphics[width=0.5\linewidth ]{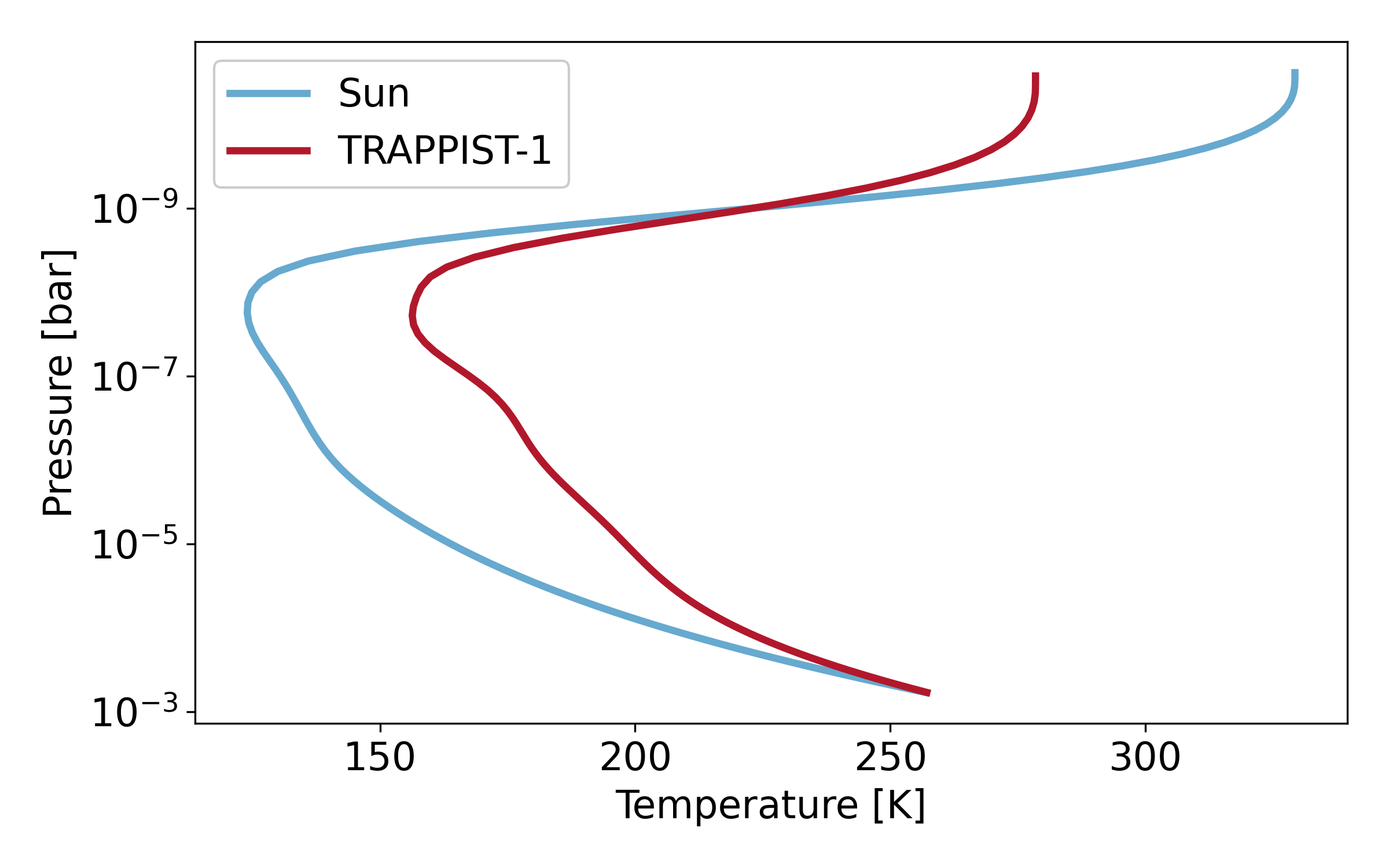}~~~   \includegraphics[width=0.5\linewidth ]{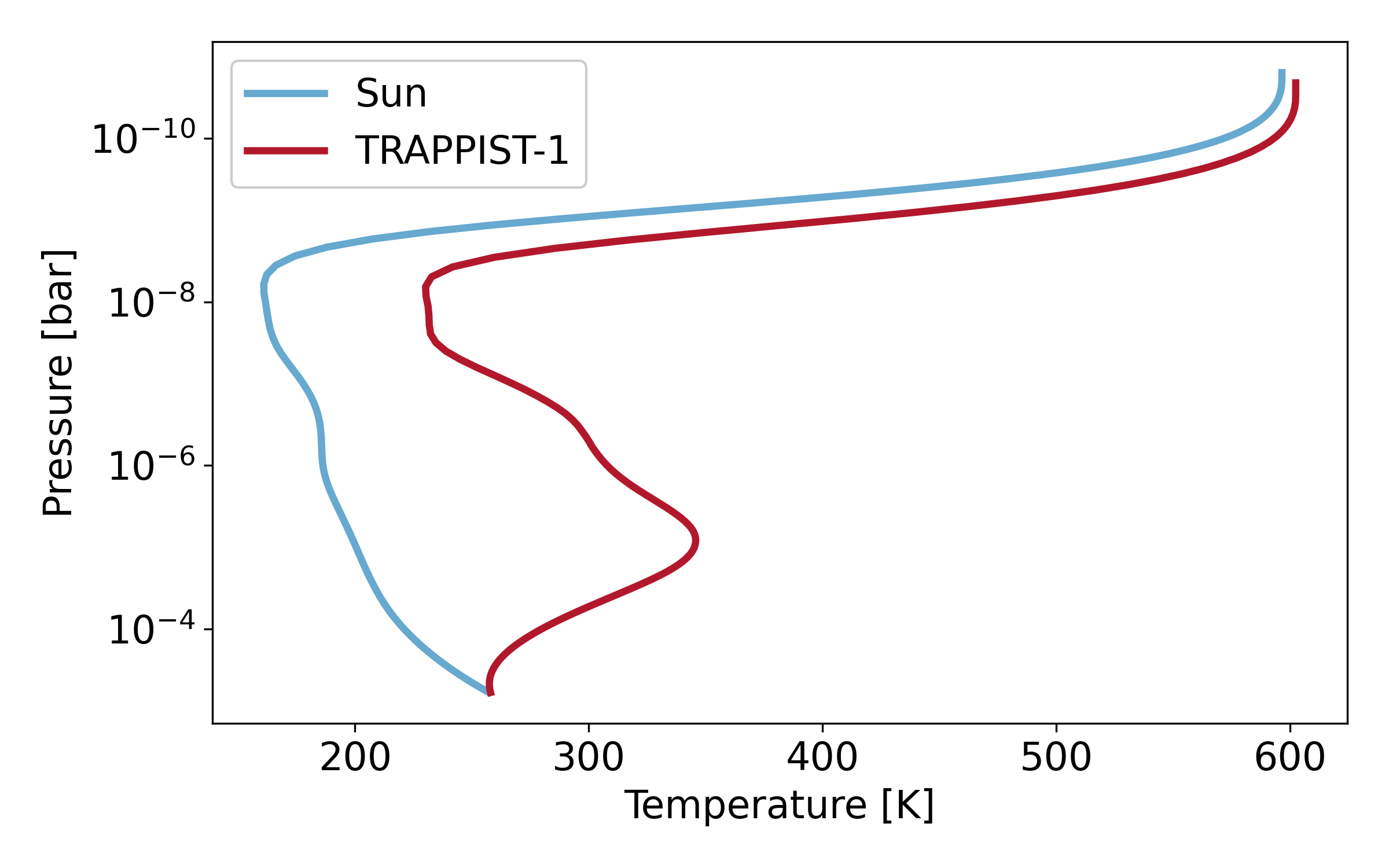}

   \caption{Pressure-temperature profiles for a planet with a 1\%~N$_2$-99\%~CO$_2$ exposed to the solar spectrum (in blue) and the TRAPPIST-1 spectrum (in red). Left: Receiving 1 F$_{{\rm EUV,}\oplus}$. Right: Receiving 4 F$_{{\rm EUV,}\oplus}$.}
   \label{fig:specvar}%
   \end{figure*}

To test the impact of these differences in the stellar XUV spectra, we also calculated several models using the Mega-MUSCLES spectrum. The pressure-temperature profiles in Figure \ref{fig:specvar} show how a 1~M$_{\oplus}$ planet with a 1\%~N$_2$-99\%~CO$_2$ composition reacts to the solar spectrum and the TRAPPIST-1 spectrum. Left in Fig. \ref{fig:specvar}, the modelled planet receives 1~F$_{{\rm EUV,}\oplus}$, which corresponds to a distance of 1~au in the Solar System and 0.512~au in the TRAPPIST-1 system. The right panel of Fig. \ref{fig:specvar} represents a planet receiving 4~F$_{{\rm EUV,}\oplus}$, which corresponds to a distance of 0.5~au for the Solar System and 0.256~au for the TRAPPIST-1 system.

These figures demonstrate that the details of the spectrum can strongly influence the temperature profile of an atmosphere. They show that the same upper atmosphere is colder at most altitudes irradiated by the solar spectrum compared to the synthetic TRAPPIST-1 spectrum, even though, as shown in Fig. \ref{fig:spectrum}, the atmosphere receives more total energy due to the stronger photospheric radiation. We ran models excluding the UV with $\lambda$ > 121 nm and X-ray ($\lambda$ < 10 nm) sections of the spectrum, which resulted in similar pressure-temperature profiles. This indicates that the shape of the temperature profile is dominated by the EUV (10 nm < $\lambda$ < 121 nm) section of the spectrum. However, even though the flux integrated over this range is equal for both spectra, the strengths in individual emission lines in this range can influence the thermal structure. Preliminary investigations of the spectra and chemical compositions of these models seem to indicate that these particular differences are caused by a peak in the cross section of the dissociation reaction of CO$_2$ into CO and O. This peak occurs around 109~nm, where the TRAPPIST-1 modelled flux is almost two orders of magnitude below the observed solar flux. This region of lower flux also corresponds to the point at which the Mega-MUSCLES spectrum changes from the Phoenix model to the differential emission measure. As described before, this section of the spectrum is strongly affected by interstellar extinction and leaves great uncertainty in the modelled spectra. We therefore used the solar spectrum in this work. A more detailed analysis of all of the particular wavelengths that influence the temperature profile is beyond the scope of this paper.

\subsection{Atmospheric survival zone}

   \begin{figure*}
   \sidecaption
    \includegraphics[width=12cm]{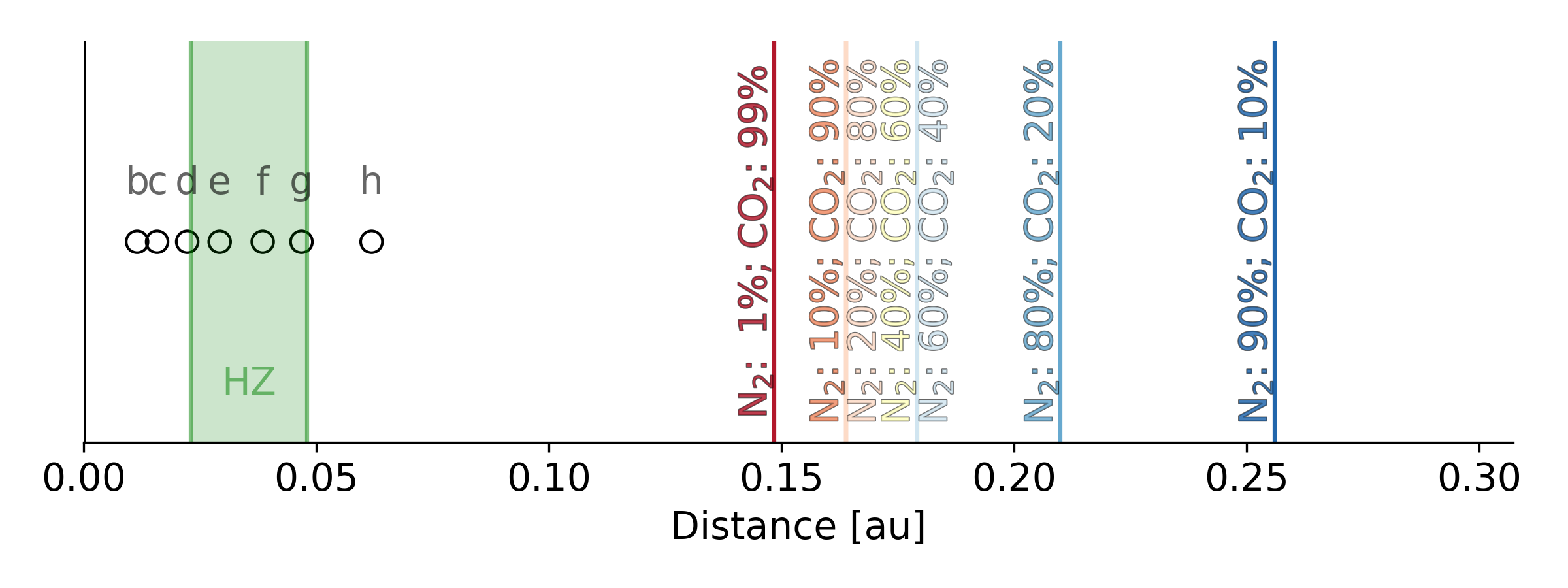}
        \caption{Overview of the planets in the TRAPPIST-1 system and the estimated habitable zone (indicated by the green lines, taken from \citealt{bolmont2017}). We added vertical lines at the minimum distances at which atmospheres of various compositions could survive for more than 1 Gyr.}
        \label{fig:hzaz}%
    \end{figure*}
    
Throughout this work, we used the TRAPPIST-1 system as an example to derive representative values for other, real exoplanet systems. However, by modelling a grid around these parameters, we kept our approach as general as reasonable so that it can be applied more easily to host stars with rocky planets other than, but similar to, TRAPPIST-1. Figure \ref{fig:hzaz} illustrates that the habitable zone of the TRAPPIST-1 system does not overlap with the region in which an atmosphere could survive for more that 1 Gyr. As discussed in Sec. \ref{sec:evol}, this can be attributed to the extended periods of high X-ray to bolometric luminosity that fully convective stars exhibit. To draw conclusions about not fully convective M dwarfs or other spectral types, Fig. \ref{fig:hzaz} should be extended to different stars with different ratios of the X-ray to bolometric luminosity. This could give insights into the type of star for which a planet in the habitable zone  would be expected to be able to hold on to an atmosphere. As demonstrated by Fig. \ref{fig:massvar}, the mass of the planet can significantly influence the mass-loss rate as well. To properly extend our conclusions to other planetary systems, a wider range of planet masses should also be modelled.

\section{Conclusions}
\label{sec:conc}
We made use of a self-consistent thermo-chemical upper-atmosphere code to calculate the thermal losses for various exoplanet scenarios, considering rocky planets with N$_2$-CO$_2$ atmospheres. The investigated parameter space covers the majority of the planets in the TRAPPIST-1 system. This system was chosen owing to its variety of rocky planets and because the scrutiny with major telescopes such as JWST is ongoing. Using the grid, we explored several parameters (received irradiance, planetary mass, and atmosphere composition) and determined their associated Jeans escape to assess the conditions that would favour the long-term survival of an atmosphere. Predictably, the Jeans loss increases as the irradiance increases or the planetary mass decreases. As demonstrated in \citet{johnstone2021}, we find that the Jeans loss decreases as the mixing ratio of CO$_2$ increases because CO$_2$ is an effective coolant of upper atmospheres.

Simulating these different effects with the parameters of the TRAPPIST-1 system, we conclude that all planets could lose one bar of atmosphere in 1 Myr, regardless of the mixing ratio between N$_2$ and CO$_2$. When we take the early evolutionary stages of the star into account, the planets would have been exposed to even higher irradiance levels because very low-mass dwarfs spin down much more slowly than Sun-like stars, and their deeper convection zones drive a dynamo. This means that their saturated phase at maximum XUV activity lasts up to several \textnormal{billion years}. This has been illustrated in particular for the TRAPPIST-1 system by \citet{wheatley2017}, \citet{fleming2020}, and \citet{birky2021}.

Furthermore, we attempted to keep our models as general as possible for M-type dwarfs through the use of a grid of models rather than tailor-made models. This allowed us to apply our method to other planetary systems. The results of our models tentatively indicate that the habitable zone of M dwarfs after their arrival on the main sequence is not suited for the long-term survival of secondary atmospheres around planets of the considered planetary masses owing to the high ratio of spectral irradiance of XUV to optical/infrared radiation over a very long time compared to more massive stars. Maintaining atmospheres on planets like this requires their continual replenishment or their formation very late in the evolution of the planets. A further expansion of the grid and more detailed studies of the parameter space are required to draw definitive conclusions for the entire spectral class of M dwarfs.

Our conclusion from this work is therefore significant for terrestrial planets with a mass that is similar to the Earth's mass that orbit mid- to late-M dwarfs such as TRAPPIST-1 near or inside the (final) habitable zone. For these planets, substantial N$_2$/CO$_2$ atmospheres are unlikely unless atmospheric gas is continually replenished at high rates on timescales of no more than a few \textnormal{million years} (the loss timescales estimated in our work), for example, through volcanism.

\begin{acknowledgements}
      The authors would like to thank the anonymous reviewer for their in depth review of this work and the detailed comments they provided.
      The results of this work were partially achieved at the Vienna Scientific Cluster (VSC).
      We acknowledge the Community Coordinated Modeling Center (CCMC) at Goddard Space Flight Center for the use of the Instant Run System of the NRLMSIS model, https://kauai.ccmc.gsfc.nasa.gov/instantrun/nrlmsis/.
\end{acknowledgements}

%
%
\bibliographystyle{aa}
\bibliography{citations.bib}

\begin{appendix} 
   \begin{figure*}
\section{Additional figures}
\label{app1}
   \centering
   \includegraphics[width=0.9\linewidth ]{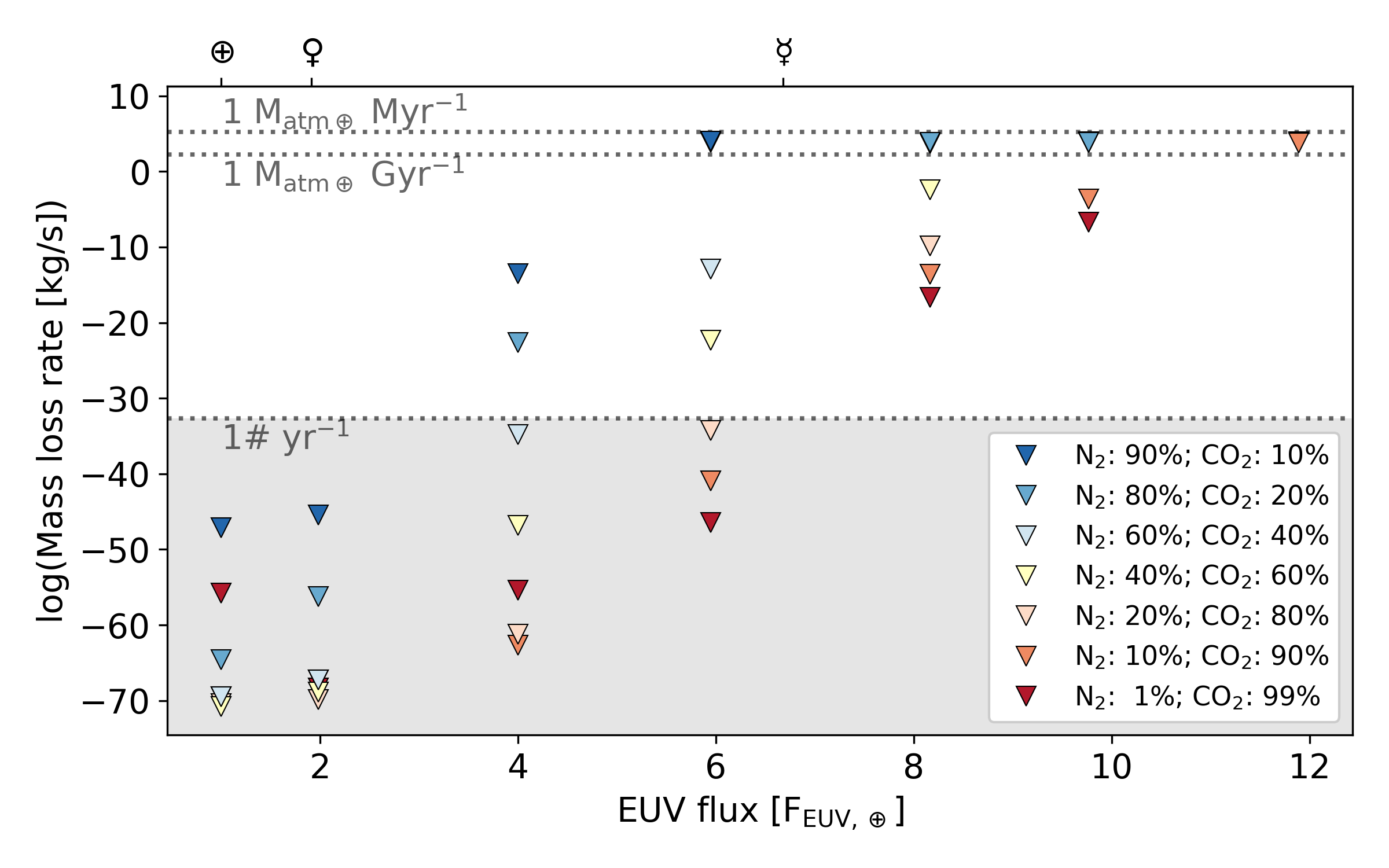}
   \caption{Abundance-weighted average Jeans escape for a 0.8~ M$_{\oplus}$ planet. The bulk compositions are coloured consistently with Fig. \ref{fig:m10_full}. The ticks on the top x-axis indicate the EUV flux received by the Solar System planets (given by their astronomical symbols). The horizontal lines indicate mass-loss rates in different units. The grey area indicates negligible loss rates.}
    \label{fig:m08_full}%
    \end{figure*}

   \begin{figure*}
   \centering
   \includegraphics[width=0.9\linewidth ]{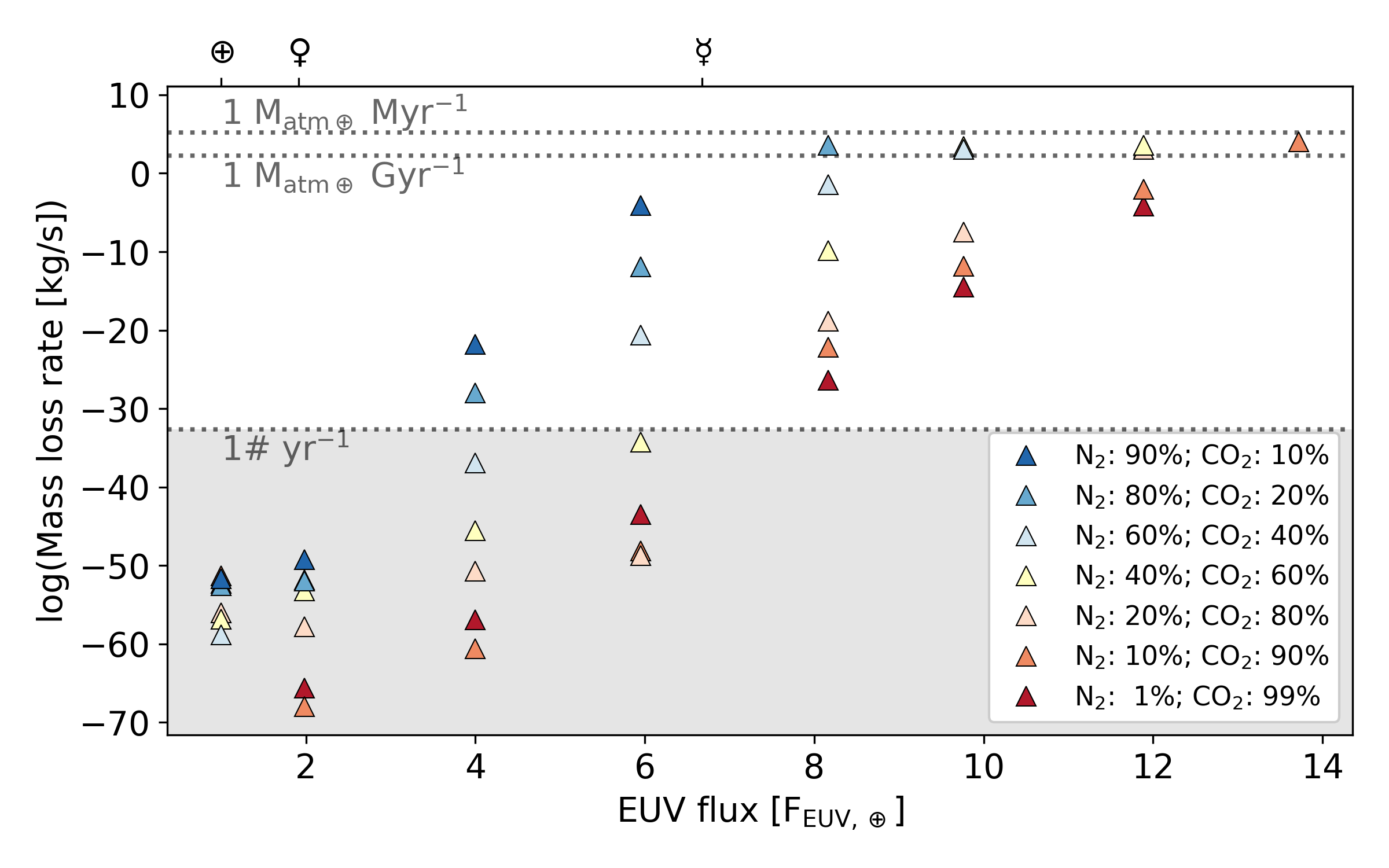}
   \caption{Abundance-weighted average Jeans escape for a 1.2~M$_{\oplus}$ planet. The bulk compositions are coloured consistently with Fig. \ref{fig:m10_full}. The ticks on the top x-axis indicate the EUV flux received by the Solar System planets (given by their astronomical symbols). The horizontal lines indicate mass-loss rates in different units, and the grey area indicate negligible loss rates.}
    \label{fig:m12_full}%
    \end{figure*}

   \begin{figure*}
   \centering
   \includegraphics[width=0.9\linewidth ]{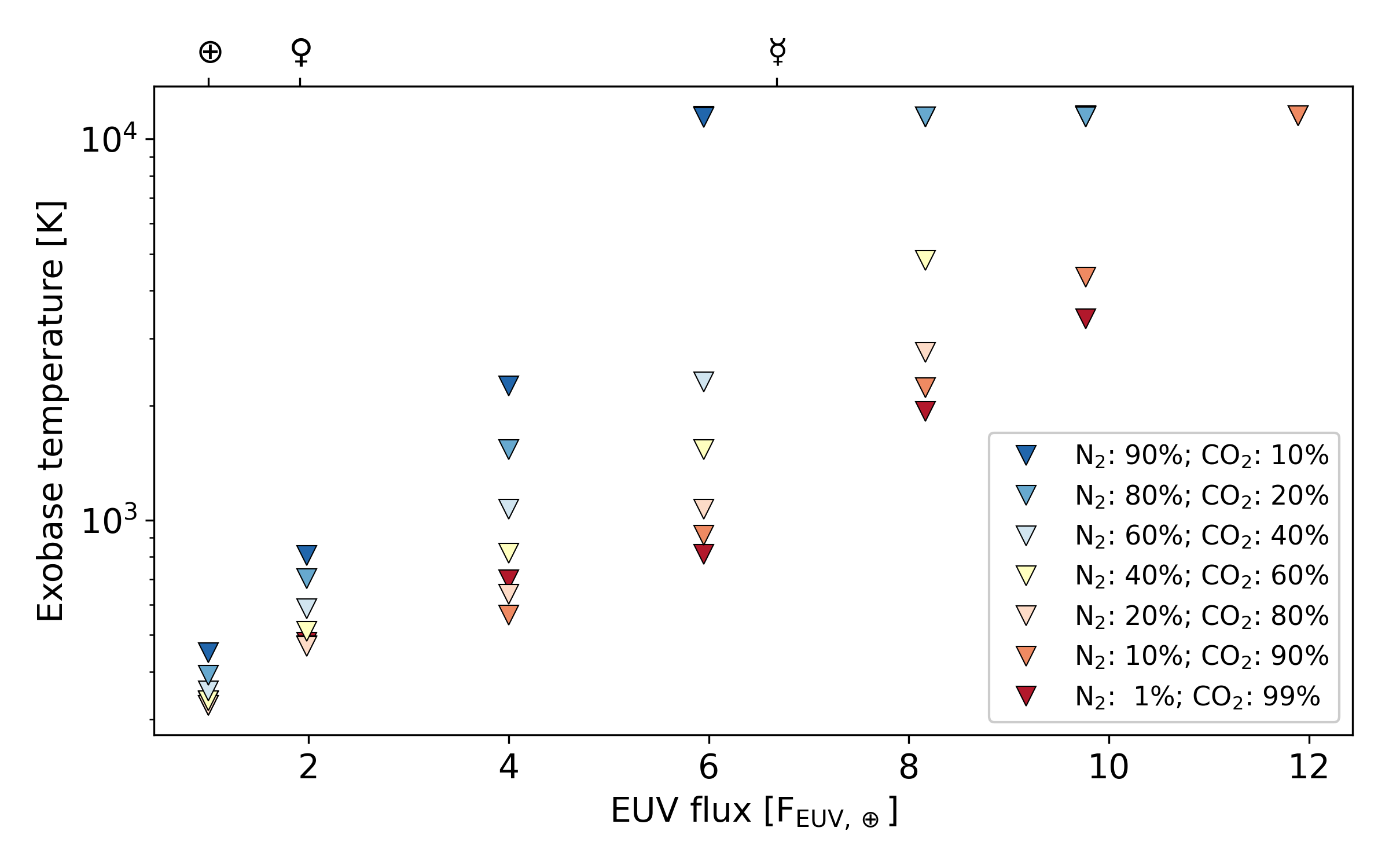}
   \caption{Exobase temperature for a 0.8~M$_{\oplus}$ planet. The bulk compositions are coloured consistently with Fig. \ref{fig:m10_full}. The ticks on the top x-axis indicate the EUV flux received by the solar system planets (given by astronomical symbol).}
    \label{fig:temp_0p8}%
    \end{figure*}

   \begin{figure*}
   \centering
   \includegraphics[width=0.9\linewidth ]{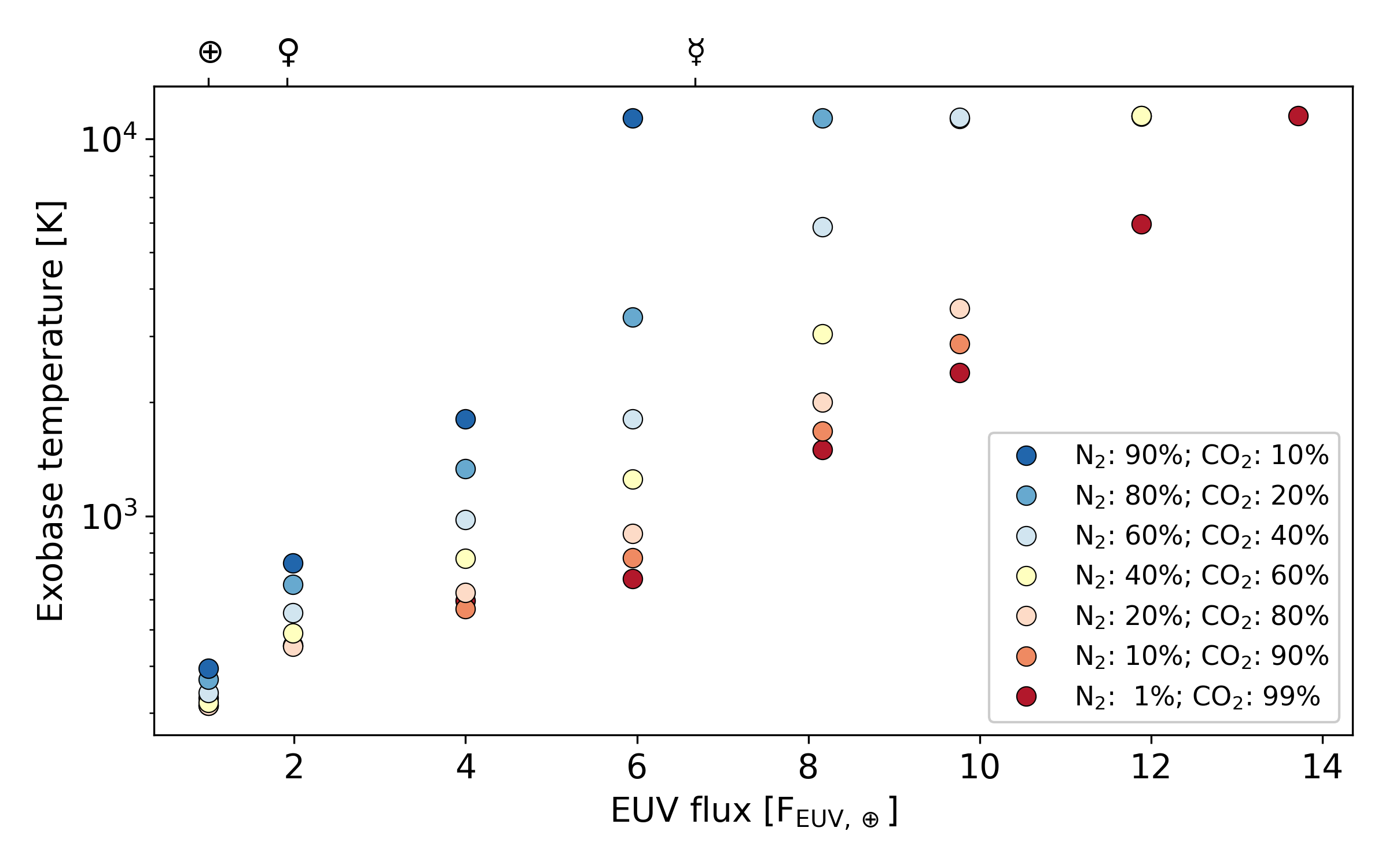}
   \caption{Exobase temperature for a 1.0~M$_{\oplus}$ planet. The bulk compositions are coloured consistently with Fig. \ref{fig:m10_full}. The ticks on the top x-axis indicate the EUV flux received by the Solar System planets (given by their astronomical symbols).}
    \label{fig:temp_1p0}%
    \end{figure*}

   \begin{figure*}
   \centering
   \includegraphics[width=0.9\linewidth ]{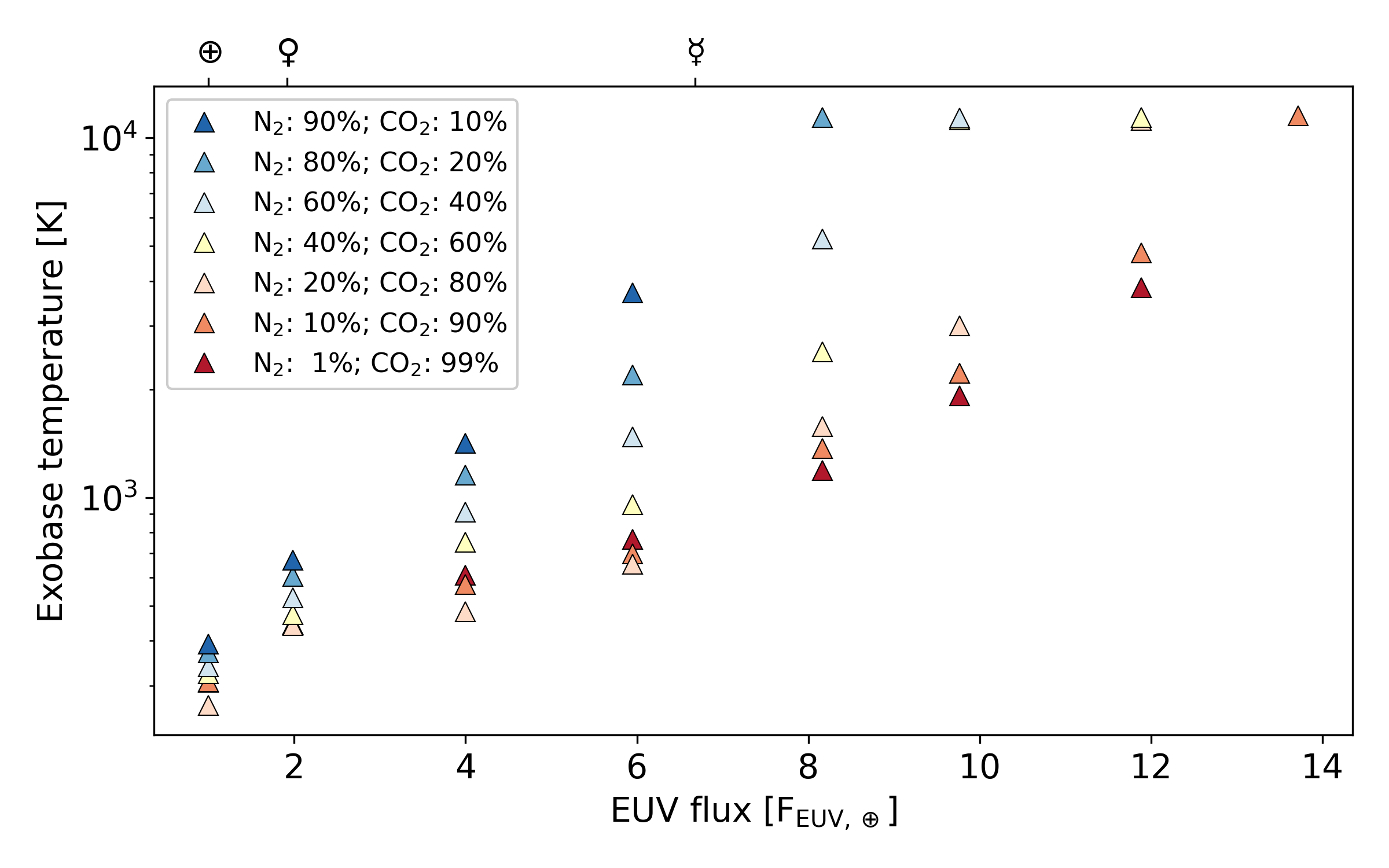}
   \caption{Exobase temperature for a 1.2~M$_{\oplus}$ planet. The bulk compositions are coloured consistently with Fig. \ref{fig:m10_full}. The ticks on the top x-axis indicate the EUV flux received by the Solar System planets (given by their astronomical symbols).}
    \label{fig:temp_1p2}%
    \end{figure*}

   \begin{figure*}
   \centering
   \includegraphics[width=0.9\linewidth ]{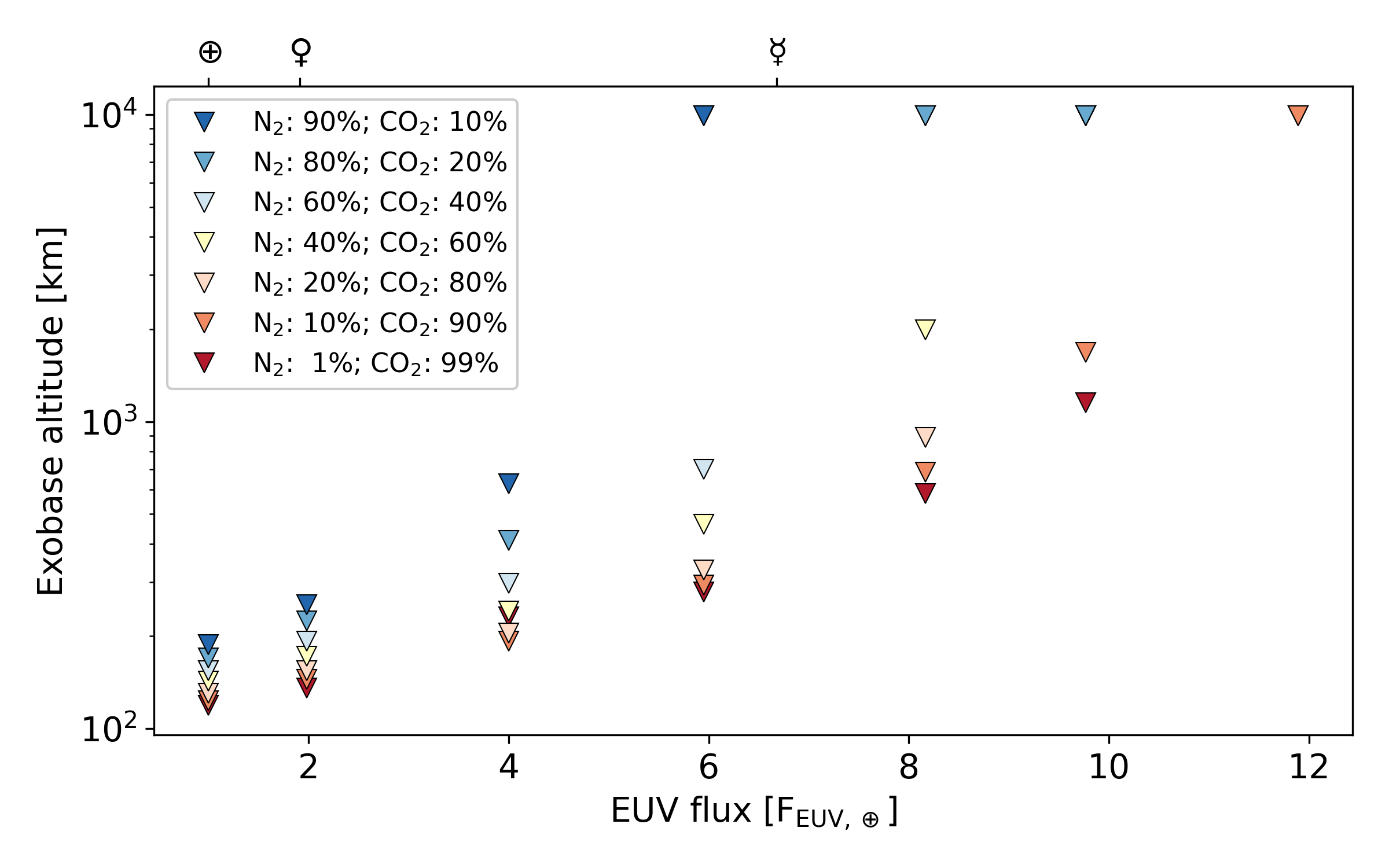}
   \caption{Exobase altitude for a 0.8~M$_{\oplus}$ planet. The bulk compositions are coloured consistently with Fig. \ref{fig:m10_full}. The ticks on the top x-axis indicate the EUV flux received by the Solar System planets (given by their astronomical symbols).}
    \label{fig:alt_0p8}%
    \end{figure*}

   \begin{figure*}
   \centering
   \includegraphics[width=0.9\linewidth ]{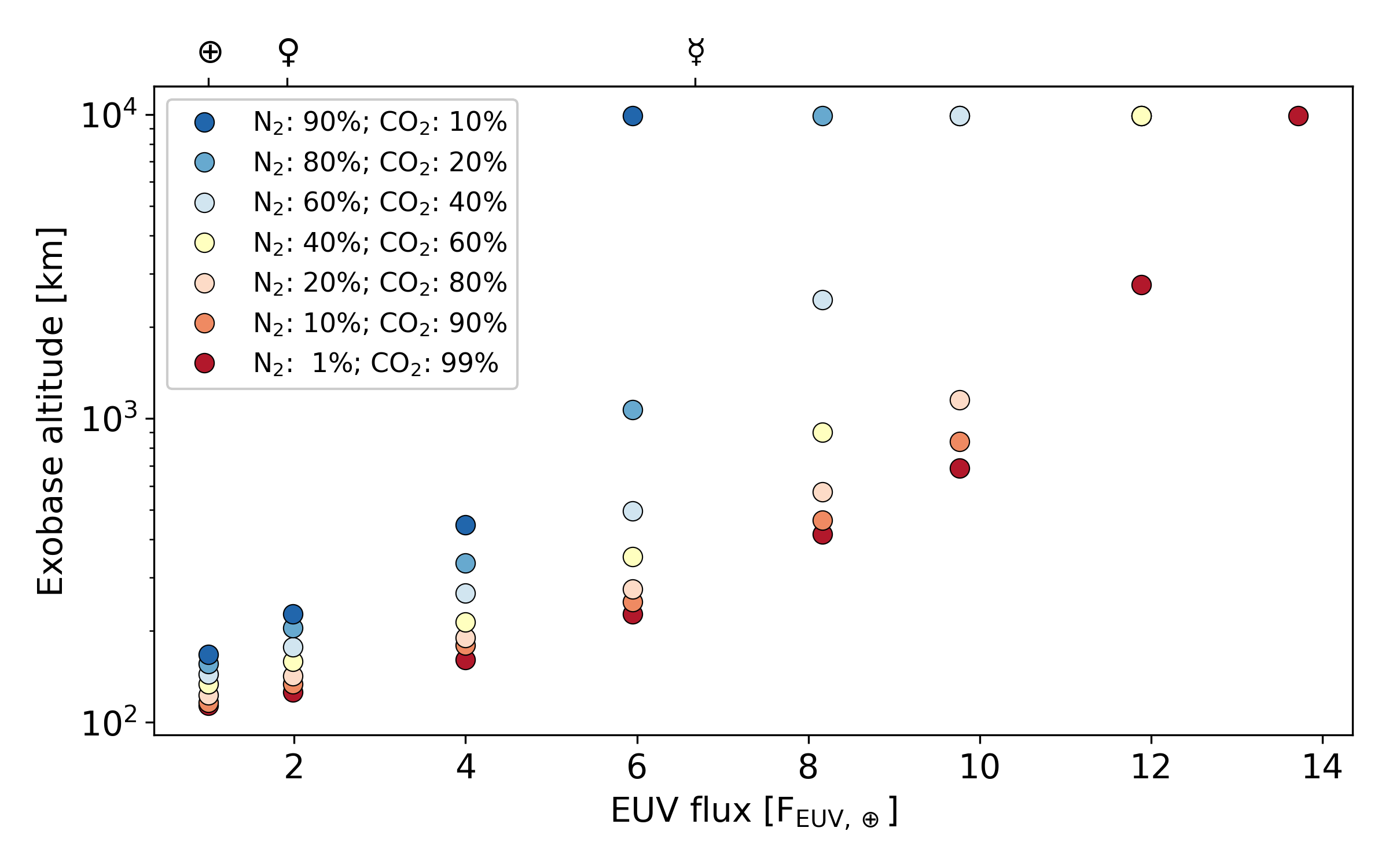}
   \caption{Exobase altitude for a 1.0~M$_{\oplus}$ planet. The bulk compositions are coloured consistently with Fig. \ref{fig:m10_full}. The ticks on the top x-axis indicate the EUV flux received by the Solar System planets (given by their astronomical symbols).}
    \label{fig:alt_1p0}%
    \end{figure*}

   \begin{figure*}
   \centering
   \includegraphics[width=0.9\linewidth ]{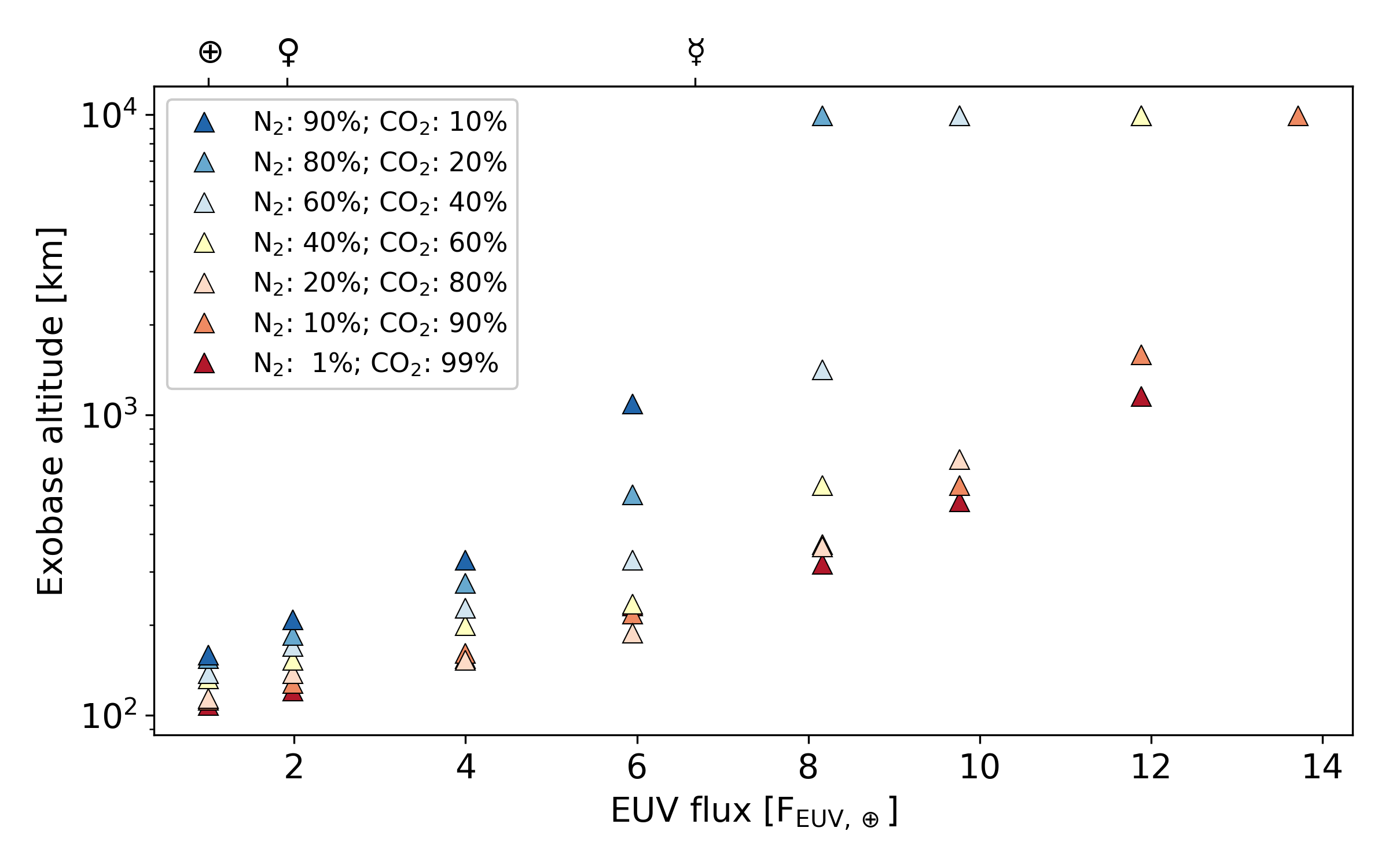}
   \caption{The exobase altitude for a 1.2~M$_{\oplus}$ planet. The bulk compositions are coloured consistently with Fig. \ref{fig:m10_full}. The ticks on the top x-axis indicate the EUV flux received by the Solar System planets (given by their astronomical symbols).}
    \label{fig:alt_1p2}%
    \end{figure*}

\end{appendix}
\end{document}